\documentclass[11pt,a4paper]{article}
\usepackage{bm}
\usepackage{xspace}
\usepackage{geometry}
\usepackage{array}
\usepackage{subfigure}
\usepackage{epsfig}
\usepackage{hhline}
\usepackage{natbib}
\usepackage{bbm}
 \usepackage[T1]{fontenc}
 \usepackage[]{graphicx,float,latexsym,times}
 \usepackage{amsfonts,amstext,amsmath,amssymb,amsthm}
 
 \RequirePackage[OT1]{fontenc}
\usepackage[utf8]{inputenc}    
\DeclareMathAccent{\widehat}{\mathord}{largesymbols}{"62}
\DeclareMathAccent{\widetilde}{\mathord}{largesymbols}{"65}

\def\ebo{\textrm{\mathversion{bold}$\mathbf{\beta}^0$\mathversion{normal}}}

\def\oo{\textrm{\mathversion{bold}$\mathbf{0}$\mathversion{normal}}}
\def\eb{\textrm{\mathversion{bold}$\mathbf{\beta}$\mathversion{normal}}}  
 
\def\el{\textrm{\mathversion{bold}$\mathbf{\lambda}$\mathversion{normal}}}

\def\eR{I\!\!R}
\def\eE{\mathbb{E}}
\def\eP{I\!\!P}
\def\e1{1\!\!1}

\newtheorem{theorem}{Theorem}[section]
\newtheorem{corollary}{Corollary}[section]
\newtheorem{lemma}{Lemma}[section]
\newtheorem{remark}{Remark}[section]

\newcommand{\beqn}{\begin{eqnarray*}}
\newcommand{\eeqn}{\end{eqnarray*}}
  
\def\ee1{\textrm{\mathversion{bold}$\mathbf{\varepsilon}$\mathversion{normal}}}  
\def\eth{\textrm{\mathversion{bold}$\mathbf{\theta}$\mathversion{normal}}}  

\def\oo{\textrm{\mathversion{bold}$\mathbf{0}$\mathversion{normal}}}

\def\eu{\mathbf{{u}}}
\def\eg{\mathbf{{g}}}
\newcommand{\N}{\mathbb{N}}
\newcommand{\R}{\mathbb{R}}
\newcommand{\PP}{\mathbb{P}}

\def\eX{\mathbf{X}}
\def\ex{\mathbf{x}}

\newcommand{\Var}{\mathbb{V}\mbox{ar}\,}

\def\argmin{\mathop{\mathrm{arg\,min}}} 

\begin{document}
 
\title {{\bf Smoothed empirical likelihood estimation and automatic variable selection for an expectile high-dimensional model with possibly missing response variable }}
 
    \date{}
   \maketitle
  \author{
  	\begin{center}
  		  Gabriela CIUPERCA \\
  	  		\small{Université Claude Bernard Lyon 1, UMR 5208, Institut Camille Jordan, \\
  	  			Bat.  Braconnier, 43, blvd du 11 novembre 1918, F - 69622 Villeurbanne Cedex, France.
  		}
  	\end{center}
  }
   
 \begin{abstract}
  We consider a linear model which can have a large number of explanatory variables, the errors with an asymmetric distribution or some values of the explained variable are missing at random. In order to take in account these several situations, we consider the non parametric empirical likelihood (EL) estimation method.  Because a constraint  in EL contains an indicator function then a smoothed function instead of the indicator will be considered. Two  smoothed expectile  maximum EL methods are proposed, one of which will automatically select the explanatory variables. For each of the methods we obtain the  convergence rate of the estimators and their asymptotic normality. The smoothed expectile empirical log-likelihood ratio process follow asymptotically a chi-square distribution and moreover the adaptive LASSO smoothed expectile maximum EL estimator satisfies the sparsity property which guarantees the automatic selection of zero model coefficients. In order to implement these methods, we propose four algorithms.
 \end{abstract}
  
\section{Introduction}
The empirical likelihood (EL) estimation method introduced by \cite{Owen.90} is a non parametric statistical technique. The advantages of the EL method for a parametric regression model  are the following. First, this method allows to give the  acceptance zone for hypothesis tests on the multidimensional parameter and it also allows to find the confidence interval for model parameter.   Thus, the obtained theoretical results allow to find a simple test statistic, useful in applications.  Moreover, if the model includes outliers then the EL method allows to obtain a  most accurate and robust prediction. \\
On the other hand, as is often the case in applications, a model may contain missing data for the response variable, the explanatory variables may be very large in number, or the model errors may have an asymmetric distribution. All these cases will be considered in the present paper using the EL technique.  For taking into account the asymmetry of the error distribution, the constraint in the EL process will be based on the expectile function which contains an indicator function, only that the non differentiability of the constraint poses problems in the theoretical study of EL. Then, in order to regularize the constraint  we replace the indicator function with a smoothed function constructed with   a kernel density.  If the model is high-dimensional, it will also be necessary to carry out an automatic selection of the relevant explanatory variables. To do this, we add an adaptive LASSO-type penalty to the empirical log-likelihood process. To highlight the contribution of this paper, we give references to the bibliography on the subject. 
\cite{Chen-Mao.21} considers a generalized linear model in high-dimension, without missing data, where the parameters are estimated by the empirical maximum likelihood penalized by an adaptive LASSO penalty. They show that the estimator obtained satisfies the oracle properties, including the automatic selection of null parameters and that the empirical log-likelihood ratio has an asymptotic $\chi^2$ distribution. \cite{Ren-Zhang.11}  considers a SCAD penalty for empirical log-likelihood. The estimator obtained satisfies also  the oracle properties. An algorithm based on a BIC type criterion makes possible to find the tuning parameter of the penalty. In \cite{Tang-Leng.10}  the number of parameters of the linear model depends on the number $n$ of observations and   the associated constrain  corresponds to the least squares method. The same SCAD penalty is considered in  \cite{Zhao-Haziza-Wu.22} for a more general case of estimating function. Note that in \cite{Liu:Zou:Wang:13} a constant model of dimension $p$ depending on $n$ is considered however only the mean of the random vector is estimated by the EL method. For a linear model, in  \cite{Guo:Zou:Wang:Chen:13}, the design can be either deterministic or random  where the associated constraint in the EL process is those corresponding to the least squares (LS). The asymptotic distribution of the  empirical log-likelihood ratio is  chi-square  when the design is random and Gaussian when the design is deterministic.  \cite{Qin.Li.Lei.09} also considers a linear model with a deterministic design and with missing data for the explained variable. The proposed method reconstructs the missing data and then the asymptotic distribution of the reconstituted empirical log-likelihood ratio is chi-square.\\
For a quantile linear regression with the explained variable  missing at random (MAR),  \cite{Zhang-Wang.20}  estimates the model parameters by the EL method. Moreover, the SCAD penalty also allows automatic selection of variables. Still for a quantile linear model with missing covariables of type MAR, \cite{Liu-Yuan.2016} estimates and studies the asymptotic properties of a weighted EL estimator.\\
Using a constraint function that does not satisfy the conditions considered in the present paper, how we will see later, \cite{Ozdemir-Arslan.2021} considers a linear model without missing data where the model parameters are estimated by the EL method without penalty.  The estimator obtained is asymptotically Gaussian\\
For a generalized linear model with weakly dependent high-dimensional data,  \cite{Zhang-Shi-Tian-Xiao.19}  proves that the EL estimator with a penalty as in  \cite{Fan-Li.01} satisfies the oracle properties even if the observations are dependent.  \cite{Zhang-Shi-Tian-Xiao.19}   generalizes the results of \cite{Leng-Tang.12} where i.i.d. random vectors with a distribution dependent on a vector of unknown parameters is considered.
Still for the framework in a high-dimensional model, \cite{Chang-Tang-Wu.18} proposes two  penalties for the empirical  log-likelihood ratio for doing variable selection.\\
In all these references, to the author's knowledge, the case of missing data for the explained variable, for an expectile high-dimensional model, has not been considered. Then, in the present paper, two  smoothed expectile  maximum EL methods are proposed. A first method useful especially when the model  with few insignificant variables is not of high-dimension and a second method useful for high-dimensional models, which will automatically select the explanatory variables. The asymptotic behavior of the estimators is studied for each of the methods, more precisely we obtain the  convergence rate of the estimators and their asymptotic normality.  We also show that the smoothed expectile empirical log-likelihood ratio process follow asymptotically a chi-square distribution and that the adaptive LASSO smoothed expectile maximum EL estimator satisfies the sparsity property which guarantees the automatic selection of zero coefficients. In order to implement these methods, we propose four algorithms,  two for each type of estimator. These algorithms and corresponding estimation methods will be validated firstly by numerical simulations and afterwards by an application on real data.\\
The paper is organized as follows. In Section \ref{sect_model} we introduce the model, the principle of the expectile empirical likelihood method, general notations and common assumptions for the two estimation methods. Section \ref{section_EMVexpectile} defines and studies the smoothed expectile empirical likelihood process and estimator. Section \ref{section_EMVaLASSO} is dedicated to the estimator and process which will allow the automatic selection of the explanatory variables of a high-dimensional model. Section \ref{sect_alg} presents the algorithms for calculating the two estimators. In Section \ref{section_simu}, simulations are firstly carried out, followed by an application on real data presented in Section \ref{sect_appli}. The theoretical result proofs are relegated in Section \ref{sect_proof}.
\section{Notations, model and suppositions}
\label{sect_model}
Let us start this section with some notation that will be used throughout the paper. 
Note that all vectors are considered column, moreover, matrices and vectors are denoted by boldface uppercase and lowercase  letters. For a vector or matrix, the symbol $T$ at the top right is used for their transpose. We denote by $\mu_{\min}$ and $\mu_{\max}$  the smallest and largest eigenvalue of a positive definite matrix, respectively, by $\|.\|$  the Euclidian norm of a vector and by $\textbf{0}_q$ the   $q$-vector  with all components 0. For an event $E$, $\e1_E$ denotes the indicator function that the event $E$ happens.  Throughout this paper, we use $C$ to denote a positive generic constant, without interest, which does nor depend on $n$. If $(a_n)_{(n \in \N)}$ and $(b_n)_{(n \in \N)}$ are two positive deterministic sequences such that $\underset{n \rightarrow \infty}{\text{lim}} a_n/b_n=0$, then we denote this by $a_n=o(b_n)$ or by $a_n \ll b_n$. We will also use the following notations: if $U_n$ and $V_n$ are two random variable sequences, notation $V_n=o_\PP(U_n)$ means that $\underset{n \rightarrow \infty}{\text{lim}} \mathbb{P}(|  {U_n}/{V_n} | >e)=0 $ for all $ e >0$ .  Moreover, notation $V_n=O_\PP(U_n)$ means that there exists $ C >0 \, \text{so that} \, \underset{n \rightarrow \infty}{\text{lim}} \mathbb{P}(|  {U_n}/{V_n} | >C )<e  $ for all $ e>0$. For an index set ${\cal A}$, a parameter vector $\eb$, a square matrix $\textbf{M}$, we denote by $\eb_{\cal A}$  the sub-vector of $\eb$ which contains the components $\beta_j$ with $j \in {\cal A}$ and by $\textbf{M}_{\cal A}$ the sub-matrix of $\textbf{M}$ with row and column indexes in ${\cal A}$. The cardinality of ${\cal A}$ is denoted by $|{\cal A}|$. For a vector $\textbf{v}$, $\textrm{diag}(\textbf{v})$ denotes the diagonal matrix with the elements of $\textbf{v}$ on the diagonal. For real $x$ we use the notation $\textrm{sgn}(x)$ for the sign function $\textrm{sgn}(x)={x}/{|x|}$ when $x \neq 0$ and $\textrm{sgn}(0)=0$ and for a function $F$ of scalar argument we denote its derivative by $F'$.   \\

\noindent We consider a classical linear model on $n$ observations:
\begin{equation}
\label{eq1_1}
Y_i=\eX_i^\top \eb+\varepsilon_i, \qquad i=1, \cdots , n, 
\end{equation}
with $Y_i$ the response variable, $\eb \in \R^p$  the vector of parameters  and $\ebo$ its true value (unknown),  $\eX_i$ random vector of $p$ explanatory variables and $\varepsilon_i$  the model error. We denote by $Y$, $\eX$ and $\varepsilon$ the generic random variables (vectors) for $Y_i$, $\eX_i$ and $\varepsilon_i$. Notations $\eE_\varepsilon$ and $\eE_{\eX}$ are used to emphasize that the expectation is with respect to the distribution of $\varepsilon$ or $\eX$, respectively.\\
Compared to classical regression, in this paper we consider the possibility that the distribution of the model errors is asymmetric.  In this case, the expectile framework introduced by \cite{Newey-Powell.87} can be considered. 
For a given  expectile index $\tau \in (0,1)$, the expectile function is defined by $\rho_\tau(x)=|\tau - \e1_{x <0}| x^2$, with $x \in \R$.\\
 For the model errors $(\varepsilon_i)_{1 \leqslant i \leqslant n}$ we impose the following classical assumptions for an expectile model:
 \begin{description}
 	\item \textbf{(A1)}
 	 \begin{description}
 		\item \textbf{(a)}  	 $(\varepsilon_i)_{1 \leqslant i \leqslant n}$ are i.i.d. such that $\eE_\varepsilon[\varepsilon^4]< \infty$ and $
 \eE_\varepsilon[\varepsilon\big(\tau \e1_{\varepsilon >0}+ (1-\tau)\e1_{\varepsilon<0} \big)]=0$, that is its  $\tau$-th expectile is zero: $\eE_\varepsilon[\rho'_\tau(\varepsilon)]=0$.	
 \item \textbf{(b)} The density of  $\varepsilon$ is $f_\varepsilon$, with $f_\varepsilon$ bounded in a neighborhood of 0. The corresponding distribution function is $F_\varepsilon$.
\end{description} 
\end{description} 
Assumption (A1)(a)  is classic for the errors of an expectile model (see for example \cite{Liao.2019}, \cite{Ciuperca.21}). Assumption (A1)(b)  is also considered by \cite{Zhang-Wang.20} for a quantile regression but which imposes the following additional conditions:  $F_\varepsilon$ bounded around 0, $f_\varepsilon$ is $r$ times  continuously differentiable, with $r \geq 2$ and  $\big|f^{(s)}_\varepsilon(\eX) \big| \leq C(\eX) $, for $s=0, 2, \cdots, r$ in a neighborhood of 0, for all $\eX$. Assumption (A1)(b) is also always considered for a quantile regression by \cite{Liu-Yuan.2016} which  considers in addition the condition that $f^{(1)}_\varepsilon(.)$ exists and it is uniformly bounded. 
 \begin{remark}
 	\cite{Ozdemir-Arslan.2021} considers the empirical likelihood estimator for the parameters of a linear model, without missing data, by imposing a constraint with respect to two functions of type $ \rho_k$, $k=1,2$, which must satisfy the condition    $\rho_k(x)=\rho_k(-x)$ for all $x$, condition that is not satisfied by the expectile function. So, in the present paper  we are in a different case from the one presented in \cite{Ozdemir-Arslan.2021}.
 \end{remark}
\noindent  The response variable $Y_i$ may be missing, in exchange for the explanatory variables $\eX_i$  all the observations are measured  for $i=1, \cdots , n$.\\
  When not specified, the convergence in probability, denoted $ \overset{\PP} {\underset{n \rightarrow \infty}{\longrightarrow}}$, is with respect to the joint probability of $\eX$ and $\varepsilon$. We also denote by $\PP_\eX$ the probability law of the random vector $\eX$ and by $\PP_\varepsilon$ the probability law of the model error $\varepsilon$.\\
Let the random variable which will indicate whether the response variable is measured or not:
 \[
 \delta  \equiv \left\{
 \begin{array}{lll}
 1, & & \textrm{if } Y \textrm{ measured},\\
  0, & & \textrm{if } Y \textrm{ missing}.
 \end{array}
 \right.
 \]
 In this paper we assume that $Y_i$ is missing at random (MAR), that is: $\eP[\delta_i=1| Y_i,\eX_i]=\eP_\eX[\delta_i=1 | \eX_i]$, for any $i=1, \cdots , n$.  Thus, we denote the corresponding probability by:
\[
\pi(\ex)=\PP[\delta=1 | \eX=\ex], \qquad \textrm{for all} \quad \ex \in \R^p
.\] 
Remark that when all the observations of $Y_i$ are present then $\delta_i=1$, for $ i=1, \cdots , n$.\\
In that follows, we suppose that $\pi(\ex)$ is known, otherwise it can be estimated by a non parametric estimator as in  \cite{Ciuperca.13} or in \cite{Xue:09}.\\
With regard to $(Y_i)_{1 \leqslant i \leqslant n}$ and  $(\eX_i)_{1 \leqslant i \leqslant n}$ the following assumptions are considered: 
\begin{description}
	\item \textbf{(A2)}  $(Y_i,\eX_i,\delta_i)_{1 \leqslant i\leqslant n}$ are i.i.d. random vectors.
	\item \textbf{(A3)} Assumptions on the random design:
	\begin{description}
		   \item \textbf{(a)} $\eX=(X_1, \cdots , X_p)$ has bounded support and $\eE_\eX[\|\eX\|^4] < \infty$. 
	  \item \textbf{(b)} 		$ \underset{1 \leqslant i \leqslant n}{\text{sup}}	 \| \eX_i\| \leq C < \infty$, with probability converging to 1 as $n \rightarrow \infty$,
		 \item \textbf{(c)}  there exists two constants $C_1, C_2 >0$ such that $0 < C_1 \leq \mu_{\min} \big( \Var_\eX[\delta \eX]\big)  \leq \mu_{\max} \big( \Var_\eX[\delta \eX]\big) \leq C_2 < \infty$.
\end{description} 
\end{description} 
Assumption (A2) is also considered in \cite{Liu-Yuan.2016} where the quantile regression is estimated by the EL weighted method with the possibility that the covariables are missing. 
Assumption (A3)(a) is considered by \cite{Zhang-Wang.20}. If there is no missing data then assumption (A3)(c)  can be replaced, taking into account (A3)(a), by $n^{-1} \sum^n_{i=1} \eX_i \eX_i^\top \overset{\PP_\eX} {\underset{n \rightarrow \infty}{\longrightarrow}} \eE_\eX\big[ \eX \eX^\top\big] $ with $\eE_\eX\big[ \eX \eX^\top\big]$ a non-singular matrix,  assumption considered in \cite{Ozdemir-Arslan.2021}. Note also that in this last paper, assumption (A3)(b) is also considered.\\

\noindent Concerning the model errors  $(\varepsilon_i)_{1 \leqslant i \leqslant n}$ and the design $(\eX_i)_{1 \leqslant i \leqslant n}$ we suppose:
\begin{description}
	\item \textbf{(A4)}  $\varepsilon_i$ and $\eX_i$ are independent for any $i=1, \cdots , n$.
\end{description} 
The supposition that $\eX_i$ is independent of $\varepsilon_i $  is also found in other works for estimating the model parameters  by the   empirical likelihood method: \cite{Ozdemir-Arslan.2021}, \cite{Chen-Mao.21}. 
We emphasize that,  Assumption (A4) combined with $\eP[\delta_i=1| Y_i,\eX_i]=\pi(\eX_i)$  imply that $\delta_i$ is independent of $\varepsilon_i$.\\

\noindent If the model errors can be asymmetric,  \cite{Newey-Powell.87} introduced the expectile estimation method. Thus, in the case of a model with missing data, for a given expectile index $\tau$, the expectile estimator of $ \eb$  on the complete data is defined by:
\begin{equation}
\label{tbi}
\widetilde{\eb}_n \equiv \argmin_{\eb \in \eR^p}  \sum^n_{i=1} \delta_i \rho_\tau ( Y_i-\eX_i^\top \eb).
\end{equation}
For $\tau=1/2$, we get the LS estimator.  Note that, considering the function
\begin{equation}
\label{egi}
\eg_i(\eb) \equiv \delta_i \big(\tau + (1-2 \tau ) \e1_{Y_i -\eX_i^\top \eb <0}\big) \big(Y_i - \eX_i^\top \eb \big) \eX_i,
\end{equation}
then, $\widetilde{\eb}_n$ is the solution to the system of equations $n^{-1} \sum^n_{i=1} \eg_i(\eb)=\oo_p$. With these elements we can introduce the expectile empirical likelihood, denoted $L_n(\eb)$,  on the complete data with respect to probabilities $p_i \geq 0$ and $\eb$. The expectile EL  is the supremum of  $\prod^n_{i=1}p_i$, under the constraints for $p_i$: $\sum^n_{i=1}p_i=1$ and $\sum^{n}_{i=1} p_i  {\eg}_i(\eb) =\oo_p$. We highlight that the constraint $\sum^{n}_{i=1} p_i {\eg}_i(\eb) =\oo_p$ define the nature of the parametric estimation model. More precisely, the expectile  empirical likelihood process with respect to $\eb$ is defined by:
\begin{equation}
\label{lb}
L_n(\eb) \equiv  \sup_{0 \leq  p_i \leq 1} \bigg\{ \prod^{n}_{i=1}  p_i ; \; \sum^{n}_{i=1} p_i=1, \sum^{n}_{i=1} p_i {\eg}_i(\eb) =\oo_p  \bigg\},
\end{equation} 
which implies that the expectile empirical log-likelihood ratio is
\begin{equation}
\label{lb1}
{\cal R}_n(\eb) \equiv -2 \big( \log L_n(\eb) +n \log n\big).
\end{equation}
Let's remember that $- n \log n$ is the maximum of $\sum^n_{i=1} \log(p_i)$ only under the constraint $\sum^n_{i=1}$ $ p_i=1$. Obviously, in relation (\ref{lb}) the probabilities $p_i$ represent the probability weights given to $\eg_i(\eb)$. We will call  ${\cal R}_n(\eb) $ expectile empirical log-likelihood ratio function for $\eb$. Using idea of \cite{Owen.90}, we note that to test $\eb=\ebo$, taking into account assumption (A1)(a), comes down to testing $\eE[{\eg}_i(\ebo)]=\oo_p $, from where the associated constraint $\sum^{n}_{i=1} p_i {\eg}_i(\eb) =\oo_p$ in (\ref{lb}). \\
For the particular case $\tau=1/2$,  when $\delta_i =1$ for any $i=1, \cdots , n$ we get the EL method introduced by \cite{Owen.91} and when $Y_i$ is MAR we find the method studied by \cite{Xue:09}.  
Other papers in the literature consider a linear model with $Y$ containing missing data of the MAR type, the parameters of the model being estimated by EL method and the functions $\eg_i$ corresponding to estimation methods, other than expectile.  In \cite{Xue:09}, the function $\eg_i$ corresponds to the   least squares method and it proves that the estimators are asymptotically normal, the empirical log-likelihood ratio being asymptotically $\chi^2(p)$.  \cite{Luo-Pang.17} considered $\eg_i$ corresponding to the quantile method, obtained the asymptotic normality of the estimators and proved Wilk's theorem for the empirical log-likelihood ratio. More recently, \cite{Zhang-Wang.20} also considered the quantile regression but in addition they studied the automatic selection of the explanatory variables by penalizing the empirical log-likelihood ratio with SCAD penalty.  \cite{Liu-Yuan.2016}, \cite{Sherwood-Wang-Zhou.13} also considered quantile regressions but with missing variables. \\
In order to smooth ${\eg}_i(\eb) $, let $K$ be a kernel density with compact support on $[-1,1]$ and $G$ the corresponding distribution function. Hence, for all $x \in \R$, $K(x)=G'(x)$ and  $G(x) \equiv \int_{v <x } K(v) dv$. Moreover $G(x)=0$ for all $x \in \R \setminus [-1,1]$. For $h > 0$ we consider the function $G_h(x) \equiv G \big({x}/{h}\big)$, with $h$ the radius, for all $x \in \R$.\\
 For the kernel $K$ we suppose: 
 \begin{description}
 	\item \textbf{(A5)}   $K(x)$ and $K'(x)$ are bounded for all  $x \in [-1, 1]$.
 \end{description} 
Examples of kernels $K$ which satisfy  assumption (A5) are Epanechnikov, Quartic or Cubic.\\

\noindent The radius $h$ depends on  $n$: $h=h_n$, such that $h_n {\underset{n \rightarrow \infty}{\longrightarrow}} 0$. Other assumptions on  $(h_n)_{n \in \N}$ will be given later.  For readability reason the subscript $n$ does not appear for $h$.  \\
We introduced the functions $K$, $G$ and the radius $h$ in order to approximate the indicator function $\e1_{Y-\eX^\top \eb <0}$ by $G_h(\eX^\top \eb - Y)$ when $h \rightarrow 0$. Taking into account the expression of the function ${\eg}_i(\eb)$, consider then the following function: 
\[
\psi_h(\eX, Y; \eb) \equiv \tau + (1-2 \tau) G_h(\eX^\top \eb - Y)
\]
and then we consider  the functions:
\[
\widehat{\eg}_i(\eb) \equiv \delta_i \psi_h(\eX_i, Y_i; \eb) \big(Y_i - \eX_i^\top \eb\big) \eX_i, \qquad \textrm{for } i=1,\cdots ,n.
\]
Note that similarly to $\widetilde \eb_n$ of (\ref{tbi}), the smoothed expectile estimator $\widetilde{\eb}^{(h)}_n$ is the solution of the following system of $p$ equations: $n^{-1} \sum^{n}_{i=1} \widehat{\eg}_i(\eb)=\oo_p$.  \\
The same smoothed function $G_h(\eX^\top \eb - Y)$ for approximate the not differentiable indicator function $\e1_{Y - \eX^\top \eb <0}$ at point $\eb$ has considered by \cite{Whang.06}, \cite{Zhang-Wang.20} for quantile regression estimated by the EL method, and for the SCAD penalized EL method, respectively.
\section{Smoothed expectile EL method}
\label{section_EMVexpectile}
In this section we introduce the smoothed expectile empirical log-likelihood process from which we will have the corresponding estimators for the coefficient vector $\eb$  and for the Lagrange multipliers. Afterwards, we study their asymptotically properties.
Let $p_i$ be the associated  probability weights  to $\widehat{\eg}_i(\eb)$. Then, similarly to (\ref{lb1}), the smoothed expectile empirical log-likelihood ratio function for fixed parameter $\eb$, can be defined as:
\begin{equation}
\label{lb2}
\widehat{\cal R}_n(\eb) \equiv -2 \sup_{0 \leq  p_i \leq 1} \bigg\{ \sum^{n}_{i=1} \log (n p_i) ; \;  \sum^{n}_{i=1} p_i=1, \sum^{n}_{i=1} p_i \widehat{\eg}_i(\eb) =\oo_p  \bigg\}.
\end{equation}
The supremum in (\ref{lb2}) may be found by  the   Lagrange multipliers method  and then $\widehat{\cal R}_n(\eb)$ can be written:
\begin{equation}
	\label{Rnbl}
\widehat{\cal R}_n(\eb, \el(\eb)) =2  \sum^{n}_{i=1} \log \big( 1+\el (\eb) ^\top \widehat{\eg}_i(\eb)\big),
\end{equation}
with $\el(\eb)$, a random vector of dimension $p$,  the Lagrange multipliers vector, which is solution of the equation:
\begin{equation}
	\label{LL}
	n^{-1}  \sum^{n}_{i=1} \frac{\widehat{\eg}_i(\eb)}{1+\el(\eb)^\top  \widehat{\eg}_i(\eb)} =\oo_p.
\end{equation}
Taking into account the Lagrange multipliers, then the optimal  probabilities $p_i$, i.e. the solutions for (\ref{lb2}), are: 
\[
p_i=\frac{1}{n} \frac{1}{1+\el(\eb)^\top  \widehat{\eg}_i(\eb)}.
\]
Following idea of \cite{Qin-Law.94}, we define the smoothed   expectile maximum empirical likelihood (MEL) estimator by 
\begin{equation}
	\label{bnr}
\widehat{\eb}_n =\argmin_{\eb \in \R^p} \widehat{\cal R}_n(\eb, \el(\eb)).
\end{equation}
In order to study the properties of the estimator  $\widehat{\eb}_n$ and of the random process  $\widehat{\cal R}_n(\eb, \el(\eb))$ we must first study the properties of   $\psi_h(\eX,Y;\eb)=\tau + (1-2 \tau) G_h(\eX^\top(\eb -\ebo)-\varepsilon)$ and afterwards those of $ \widehat{\eg}_i(\eb)$.  First of all, let us give for  $G_h$  the following Lemma.\\
Note that the proofs of the results presented in this section are given in Subsection \ref{proofs_EMVexpectile}.
\begin{lemma}
	\label{Lemma_tl}
(i)	Under the two  assumptions of  (A1) we have
	\begin{equation}
		\label{eq_tl}
		G_h(-\varepsilon) =G(- {\varepsilon}/{h}) \overset{\PP_\varepsilon} {\underset{h \rightarrow 0}{\longrightarrow}} \e1_{\varepsilon < 0} .
	\end{equation}
(ii) Under assumptions (A1), (A2), (A3)(a), (A3)(b), (A4), we have for all $\eb \in \R^p$ such that $\|\eb-\ebo\| \leq C n^{-1/3}$, that
	\begin{equation}
	\label{eq_tl2}
	G_h(\eX^\top (\eb-\ebo) -\varepsilon)  \overset{\PP} {\underset{h \rightarrow 0}{\longrightarrow}} \e1_{\varepsilon < \eX^\top(\eb-\ebo)} .
\end{equation}
\end{lemma}
 
\noindent Moreover, in order to study the asymptotic behavior of  $\widehat{\eb}_n$, that is the solution of (\ref{bnr}), we also need to know the asymptotic behavior of  $\widehat{\eg}_i(\ebo)$,  studied by the following lemma.
\begin{lemma}
	\label{Lemma 1}
Under assumptions  (A1),  (A2), (A3)(a), (A3)(c), (A4), we have
	\begin{equation}
		\label{eq1}
		n^{-1/2} \sum^{n}_{i=1} \widehat{\eg}_i(\ebo) \overset{\cal L} {\underset{n \rightarrow \infty}{\longrightarrow}} {\cal N} (\oo_p,\textbf{B}),
	\end{equation}
\begin{equation}
	\label{eq2}
n^{-1} \sum^{n}_{i=1} \widehat{\eg}_i(\ebo)  \widehat{\eg}_i(\ebo)^\top  \overset{\PP} {\underset{n \rightarrow \infty}{\longrightarrow}} \textbf{B},
\end{equation}
\begin{equation}
	\label{eq4}
	\max_{1 \leqslant i \leqslant n} \| \widehat{\eg}_i(\ebo) \| =o_\PP(n^{1/2}),
\end{equation}
with the  $p$-square matrix $\textbf{B} \equiv 4^{-1} \Var_\eX[\delta \eX] \eE_\varepsilon \big[   \rho'_\tau (\varepsilon)\big]^2$.
\end{lemma}

\noindent  By the following theorem we show that the smoothed expectile MEL estimator $\widehat \eb_n$ is consistent and its   convergence rate is of order $n^{-1/3}$. We obtain that this estimator and the Lagrange multiplier $\el(\widehat{\eb}_n)$  are solutions of a system of $2p$ equations.
 
\begin{theorem}
	\label{Lemma 2.1 OA}
Under assumptions  (A1),  (A2), (A3), (A4), (A5), if  $h \rightarrow 0$ such that $n^{1/3}h \rightarrow\infty$, then the smoothed expectile  empirical log-likelihood ratio $\widehat {\cal R}_n (\eb, \el(\eb)) =2 \sum^{n}_{i=1} \log \big(1+ \el(\eb)^\top \widehat{\eg}_i(\eb) \big) $ has a minimum $\widehat{\eb}_n$ inside the ball $\| \eb -\ebo \| \leq C n^{-1/3}$ with probability converging to 1. Moreover, the estimators $\widehat{\eb}_n$  and $\widehat{\el}_n \equiv \el(\widehat{\eb}_n)$ are the solutions of the following equation systems:
	\begin{equation*}
		\left\{
		\begin{split} 
			\textbf{S}_1(\eb,\el(\eb))& \equiv  \frac{1}{n}\frac{\partial \widehat{\cal R}_n(\eb,\el(\eb))}{\partial \el} =\frac{2}{n} \sum^{n}_{i=1} \frac{\widehat{\eg}_i(\eb)}{1+ \el(\eb)^\top \widehat{\eg}_i(\eb)} =\oo_p \\
				\textbf{S}_2(\eb,\el(\eb))&  \equiv \frac{1}{n}\frac{\partial \widehat{\cal R}_n(\eb,\el(\eb))}{\partial \eb} =\frac{2}{n}\sum^{n}_{i=1} \frac{1}{1+ \el(\eb)^\top \widehat{\eg}_i(\eb)} \bigg(\frac{\partial  \widehat{\eg}_i(\eb)}{\partial \eb}\bigg)^\top  \el(\eb)=\oo_p
		\end{split}
		\right.
	\end{equation*}
and the Lagrange multipliers vector estimator $\widehat{\el}_n$ converges to $\oo_p$ with a convergence rate of order $n^{-1/3}$.
\end{theorem}
\begin{remark}
	\label{remark_3.1}
	(i) For the system of  $2p$ equations from Theorem \ref{Lemma 2.1 OA} in order to have a unique solution, it is necessary that $p<n$.\\
 (ii) From the proof of Theorem \ref{Lemma 2.1 OA}, it is important to consider the ball $\| \eb - \ebo \| \leq n^{-1/3}$  and not that of order $n^{-1/2}$ because we must show that $\widehat{\cal R}_n\big( \eb, \el(\eb)\big) \gg \widehat{\cal R}_n\big( \ebo, \el(\ebo)\big)$, what we don't happen if that's the order $n^{-1/2}$ (for $n^{-1/2}$,   $\widehat{\cal R}_n\big( \eb, \el(\eb)\big)$ and $ \widehat{\cal R}_n\big( \ebo, \el(\ebo)\big)$ have the same order of magnitude).\\
 (iii)  Similar results to those of Lemma \ref{Lemma 1} are shown in the proof of Theorem \ref{Lemma 2.1 OA}, for all $\eb$ such that $\| \eb -\ebo \| \leq C n^{-1/3}$.\\
 (iv) The condition $n^{1/3}h \rightarrow\infty$ is necessary to control the approximation of $\widehat {\cal R}_n (\eb, \el(\eb))$ by $\widehat {\cal R}_n (\ebo, \el(\ebo))$, $n^{1/3}$ being the inverse of the convergence rate of the estimator $\widehat{\eb}_n$. On the other hand, since the model in \cite{Whang.06} and \cite{Zhang-Wang.20} is quantile, the assumptions on $G$ and then on $K$ are not the same as in the present paper. As a result, the assumptions about the radius $h$ are also not the same: here we have only $h \rightarrow 0$ and $n^{-1/3}h \rightarrow \infty$, while   \cite{Whang.06}, \cite{Zhang-Wang.20} consider the conditions: $nh^{2r} \rightarrow 0$ and $nh/ \log n \rightarrow \infty$, with $r$ the order of the kernel $K$.
\end{remark}
\noindent The convergence rate of  $\widehat \eb_n$ and of $\widehat{\el}_n $  given in  Theorem \ref{Lemma 2.1 OA} can be improved by the following Theorem. Moreover, these two estimators $\widehat \eb_n$ and $\widehat \el_n$ are asymptotically Gaussian.
\begin{theorem}
	\label{Th1 ZW}
	Under the same assumptions as for Theorem \ref{Lemma 2.1 OA}, we have:\\
	(i)  $\|\widehat \eb_n -\ebo \|+\| \widehat \el_n \|=O_\PP(n^{1/2})$.\\
	(ii) $	n^{1/2} \big( \widehat \eb_n - \ebo\big)  \overset{\cal L} {\underset{n \rightarrow \infty}{\longrightarrow}} {\cal N}(\oo_p,\textbf{V}^{-1})$,  $n^{1/2} \widehat \el_n   \overset{\cal L} {\underset{n \rightarrow \infty}{\longrightarrow}} {\cal N}(\oo_p,\textbf{U}^{-1})$,\\
	with the $p$-square matrix $\textbf{V} \equiv  \eE \big[\frac{\partial \widehat{\eg}_i(\ebo) }{\partial \eb}\big]^\top  \eE \big[ \widehat{\eg}_i(\ebo) \widehat{\eg}_i(\ebo)^\top\big]  \eE \big[\frac{\partial \widehat{\eg}_i(\ebo) }{\partial \eb}\big]=  \eE \big[\frac{\partial \widehat{\eg}_i(\ebo) }{\partial \eb}\big]^\top  \textbf{B}  \eE \big[\frac{\partial \widehat{\eg}_i(\ebo) }{\partial \eb}\big]$, $\textbf{U} \equiv \textbf{B} \big( \textbf{I}_p -   \eE \big[\frac{\partial \widehat{\eg}_i(\ebo) }{\partial \eb}\big]\textbf{V} \eE \big[\frac{\partial \widehat{\eg}_i(\ebo) }{\partial \eb}\big]^\top \textbf{B}^{-1} \big)^{-1}  $. 
\end{theorem}

\noindent Note that the asymptotic normality in Theorem \ref{Th1 ZW} is in line with the results of previously studies conduced by \cite{Qin-Law.94} for the LS MEL estimator, by \cite{Zhang-Wang.20} for the smoothed weighted EL estimator and by \cite{Ozdemir-Arslan.2021} for the EL-MM estimator. Moreover, the asymptotic normality of the estimators $\widehat \eb_n$, $ \widehat \el_n$ allows hypothesis tests on parameter vectors $\eb$ and $\el$, respectively. For completing these results,  Theorem \ref{Th2 ZW} will allow us to find the asymptotic distribution of the smoothed expectile empirical log-likelihood ratio process $\widehat{\cal R}_n\big( \ebo, \el(\ebo)\big)$.
\begin{theorem}
	\label{Th2 ZW}
	Under assumptions  (A1),  (A2), (A3), (A4), (A5), we have
	\[
	\widehat{\cal R}_n\big( \ebo, \el(\ebo)\big)  \overset{\cal L} {\underset{n \rightarrow \infty}{\longrightarrow}}\chi^2(p).
	\]
	\end{theorem}

\noindent  Theorem \ref{Th2 ZW} allows both to find the asymptotic confidence region for the parameter $\eb$ and also to test  null hypothesis $H_0:  \eb=\ebo$ against  alternative hypothesis $H_1: \eb \neq \ebo$. Thus, for size $\alpha \in (0,1)$, the  acceptance zone of the null hypothesis   which also corresponds to the $(1-\alpha)$-level asymptotic confidence region is $\{\eb ; \; 	\widehat{\cal R}_n\big( \eb, \el(\eb)\big)  \leq c_{1-\alpha, p} \}$, with $c_{1-\alpha, p}$ the   $(1 -\alpha)$-level quantile of the  chi-square distribution with $p$ degrees of freedom. In order to calculate $\widehat{\cal R}_n\big( \eb, \el(\eb)\big)$ we will use the approximation given by  relation (\ref{LUb}).   In order to calculate $\widehat{\cal R}_n\big( \ebo, \el(\ebo)\big)$, its approximation given by relation (\ref{ZW2}) will be used and for $\widehat{\cal R}_n\big( \eb, \el(\eb)\big)$ that of relation (\ref{LUb}). These lead to an easier of calculation of the test statistic compared to that obtained by using the asymptotic normality of Theorem \ref{Th1 ZW}. Last but not least,  Theorem \ref{Th2 ZW} is a Wilks' theorem.
\section{Automatic selection of variables} 
\label{section_EMVaLASSO}
In this section we consider the problem of the automatic selection of the relevant explanatory variables, i.e. the variables  with non-zero coefficients. This automatic selection is very useful when the number $p$ of parameters is large. In the sequel we suppose that $p$ is smaller than $n$ for a reason explained later.\\
We denote the components of the smoothed expectile MEL estimator $\widehat \eb_n$ by $\big(\widehat \beta_{n,1}, \cdots , \widehat \beta_{n,p}\big)$, those of  the true parameter $\ebo$ by $\big(\beta^0_1, \cdots , \beta^0_p\big)$ and those  of a parameter vector  $\eb$  by $(\beta_1 , \cdots , \beta_p)$.\\
Following the same idea as for the classical adaptive LASSO estimation methods (see for example \cite{Zou.2006}, \cite{Liao.2019}, \cite{Ciuperca-16}), in order to automatically select the relevant explanatory variables we add to the smoothed empirical log-likelihood ratio   an adaptive LASSO penalty:
\[
\widehat{\cal R}^*_n(\eb, \el(\eb))\equiv 2 \sum^n_{i=1}\log \big( 1+ \el(\eb)^\top \widehat \eg_i(\eb)\big) +n \eta_n \sum^p_{j=1} \widehat \omega_{n,j} |\beta_j|,
\]
where $\eta_n \in \R_+$ is a known tuning parameter, the weight $\widehat \omega_{n,j}=|\widehat \beta_{n,j}|^{-\gamma}$, with the  power $\gamma>0$ fixed. According to Remark \ref{remark_3.1}(i), we deduce that the assumption $p <n$ is necessary to calculate the estimator $\widehat \eb_n$ which intervenes in the weights $\widehat \omega_{n,j}$.  Note also that with respect to \cite{Zhang-Wang.20}, which considers the smoothed quantile EL function with the SCAD penalty, in our constraints $\sum^n_{i=1} p_i  \widehat \eg_i(\eb)\big)= \oo_p $ the function $ \widehat \eg_i $ is obtained from the expectile function $\rho_\tau(.)$ which is derivable. These leads to advantages for the theoretical study of the estimators and also for the computational study.   Let be the index set   of the true non-zero coefficients ${\cal A} \equiv \{ j \in \{1, \cdots , p\}; \; \beta^0_j \neq 0\}$.
Without reducing the generality, we assume that  the first $q \equiv |{\cal A}|$ elements of $\ebo$ are non-zero, that is: ${\cal A}=\{1, \cdots, q\}$ and its complementary set is ${\cal A}^c=\{q+1, \cdots , p\}$.   \\  
Recall that in \cite{Tang-Leng.10}, \cite{Leng-Tang.12}, it is the SCAD penalty for LS (that is $\tau=1/2$)  EL process which is considered to automatically select the variables of a model without missing data. To the knowledge of the author, the adaptive LASSO penalty has been considered only by \cite{Chen-Mao.21} for an EL process corresponding to generalized linear models.\\

\noindent  The following assumption on $\eta_n$ and $\gamma$ is considered:
\begin{description}
	\item \textbf{(A6)}  $\eta_n {\underset{n \rightarrow \infty}{\longrightarrow}} 0$, $n^{2/3} \eta_n {\underset{n \rightarrow \infty}{\longrightarrow}} 0$ and $\gamma \in (1,3]$.
\end{description} 
The only paper (to the author's knowledge) that penalizes the empirical log-likelihood ratio with an adaptive LASSO is \cite{Chen-Mao.21}, where is considered $\gamma =1$ and the tuning parameter satisfies the assumption $n^{1/2} |{\cal A}| \eta_n {\underset{n \rightarrow \infty}{\longrightarrow}} 0$. In order to compare with the penalty of an expectile process (see \cite{Liao.2019}), the constraints on $\eta_n$ and $\gamma$ are $n^{1/2} \eta_n \eta_n {\underset{n \rightarrow \infty}{\longrightarrow}} 0$ and $n^{(\gamma +1)/2} \eta_n {\underset{n \rightarrow \infty}{\longrightarrow}} \infty$. In our case, the assumption $\gamma \in (0,3]$  will be needed to find the   convergence rate of the estimator proposed in this section, in order to control the  penalty behavior with respect to the smoothed expectile empirical log-likelihood ratio function.\\
Note also that the Lagrange multipliers vector $\el(\eb)$ intervening in $\widehat{\cal R}^*_n(\eb, \el(\eb))$ is the solution of the same equation (\ref{LL}). Moreover, as in the proof of Theorem \ref{Lemma 2.1 OA}, we show that if $\|\eb-\ebo \| \leq C n^{-1/3}$ then $\|\el(\eb)\|=O_\PP(n^{-1/3})$, which implies that  relation (\ref{L10}) remains also true.\\
On the other hand, similarly to  relation (\ref{LUb})  we have for $\eb$ such that $\| \eb -\ebo \| \leq C n^{-1/3}$, the following approximation for $ \widehat {\cal R}^*_n(\eb, \el(\eb))$:
\begin{equation}
	\label{Rob}
	\begin{split}
 \widehat {\cal R}^*_n(\eb, \el(\eb))& =n \bigg( \frac{1}{n} \sum^n_{i=1} \widehat \eg_i(\eb)\bigg)^\top \bigg( \frac{1}{n} \sum^n_{i=1}  \widehat \eg_i(\eb)  \widehat \eg_i(\eb)^\top \bigg)^{-1} \bigg( \frac{1}{n} \sum^n_{i=1} \widehat \eg_i(\eb)\bigg) \\
	& \qquad +o_\PP(n^{1/3}) +n \eta_n \sum^p_{j=1} \widehat \omega_{n,j}   |\beta_j|  .
		\end{split}
\end{equation} 
This approximation will allow us to show the following theorem.
\begin{theorem}
\label{thh}
Under assumptions (A1)-(A6), if $h \rightarrow 0$ such that $n^{1/3}h \rightarrow \infty$, then we have that  the minimum of $\widehat{\cal R}^*_n(\eb, \el(\eb))$ is realized for a parameter $\eb$ such that $\| \eb - \ebo\| \leq C n^{-1/3}$.
\end{theorem}
The proofs of the two theorems stated in this section are presented in Subsection \ref{proofs_EMVaLASSO}. 
We define then the adaptive LASSO smoothed expectile MEL estimator as the minimizer of $	\widehat{\cal R}_n^*(\eb, \el(\eb)) $:
	\begin{equation}
		\label{benr}
\widehat\eb_n^* \equiv \argmin_{\eb \in \R^p}  	\widehat{\cal R}_n^*(\eb, \el(\eb)) .
\end{equation}
According to definition of $\widehat\eb_n^*$ we denote by $\widehat \el_n^* \equiv \el(\widehat \eb_n^*)$ and the components of  $\widehat \eb_n^*$ by $\big(\widehat \beta^*_{n,1}, \cdots , \widehat \beta^*_{n,p}\big)$. In order to state the main result of this section, let us consider the index set $\widehat{\cal A}_n^* \equiv \big\{j \in \{1, \cdots, p \}; \; \widehat \beta^*_{n,j} \neq 0 \big\}$ corresponding to the non-zero coefficient estimators.
\begin{remark}
	\label{vit_el}
Since by Theorem \ref{thh} the convergence rate of $\widehat \eb_n^*$ is $O_\PP(n^{-1/3})$, using an identical approach as in the proof of Theorem \ref{Lemma 2.1 OA} we obtain that $\widehat \el_n^*$ is also $O_\PP(n^{-1/3})$. 
\end{remark}
\noindent The next theorem proves that the estimator $\widehat \eb_n^*$  satisfies the oracle properties, i.e. the sparsity and the asymptotic normality. Note that in order to guarantee the sparsity of  $\widehat\eb_n^*$, an additional condition on the power $\gamma$ is considered with respect to Theorem \ref{thh}. For the asymptotic normality of $\widehat\eb_n^*$ the conditions on $h$  considered in Theorems \ref{Lemma 2.1 OA}, \ref{Th1 ZW} or Theorem \ref{thh} are sufficient.
	\begin{theorem}
		\label{Th3 ZW}
	Under assumptions (A1), (A2), (A3), (A4), (A5), (A6), the radius $h$ such that $h \rightarrow 0$, $n^{1/3} h \rightarrow \infty$ and the power $\gamma$ such that  $n^{(\gamma +1)/3} \eta_n {\underset{n \rightarrow \infty}{\longrightarrow}} \infty$, then we have:\\
		(i) $\widehat \eb_n^*$ satisfies the sparsity property: $\underset{n \rightarrow \infty}{\lim} \PP \big[{\cal A} = \widehat {\cal A}^*_n \big] =1$.\\
		(ii) asymptotic normality: $n^{1/2} \big( \widehat \eb_n^* - \ebo\big)_{\cal A}  \overset{\cal L} {\underset{n \rightarrow \infty}{\longrightarrow}} {\cal N}\big( \oo_{|{\cal A}|}, \eE \big[ \frac{\partial \widehat \eg_i(\ebo)}{\partial \eb}\big]^\top_{\cal A}   \textbf{B}_{\cal A}   \eE \big[ \frac{\partial \widehat \eg_i(\ebo)}{\partial \eb}\big]_{\cal A} \big)$.
	\end{theorem}
\noindent We emphasize that since $n^{2/3} \eta_n {\underset{n \rightarrow \infty}{\longrightarrow}} 0$, then the supposition $n^{(\gamma +1)/3} \eta_n {\underset{n \rightarrow \infty}{\longrightarrow}} \infty$ on $\eta_n$ implies $\gamma >1$. 
From Theorems  \ref{Th2 ZW} and  \ref{Th3 ZW} we obtain the following corollary whose result is very useful for applications on real data.
 \begin{corollary}
 	\label{corrolaire_RnA} 
 		Under the same assumptions as for Theorem \ref{Th3 ZW}, we have:
 		\[
 	\widehat{\cal R}_n\big( \widehat \eb^*_{n,\widehat{\cal A}^*_n}, \el(\widehat \eb^*_{n,\widehat{\cal A}^*_n})\big)  \overset{\cal L} {\underset{n \rightarrow \infty}{\longrightarrow}}\chi^2(|_{\cal A}|).
 	\]
 \end{corollary}

\begin{remark}
  The asymptotic distribution of the Lagrange multipliers estimator $\widehat \el_n^*$ corresponding to  $ \widehat \eb_n^*$  is the same as that of $\widehat \el_n$ corresponding to $\widehat \eb_n$ given by  Theorem  \ref{Th1 ZW}(ii).
\end{remark}
\noindent Theorems  \ref{thh} and  \ref{Th3 ZW} are true for any $\eta_n$ that satisfies assumption (A6) and $n^{(\gamma +1)/3} \eta_n {\underset{n \rightarrow \infty}{\longrightarrow}} \infty$. Nevertheless, an optimal value can be founded. 
Thus, in order to find the optimal value $\widehat \eta_n^* $ for the tuning parameter $\eta_n$  we propose a Schwarz-type criterion:
\begin{equation}
	\label{crit_BIC}
	BIC(\eta_n) \equiv 	\widehat{\cal R}_n^*(\widehat\eb_n^*, \el(\widehat\eb_n^*))+\log n |\widehat{\cal A}^*_n|,
\end{equation}
with $\widehat\eb_n^*$ the estimator calculated for the parameter tuning $\eta_n$. We choose the following optimal value  $$\widehat \eta_n^*=\argmin_{\eta_n } BIC(\eta_n) ,$$ $\eta_n$ being sought among those which satisfy assumption (A6). The choice of an optimal tuning  parameter will be put into practice in Section \ref{sect_appli} for an application on real data.\\
The reader can see \cite{Zhang-Wang.20} for other types of BIC criteria  for smoothed quantile empirical log-likelihood ratio but with SCAD penalty.  A similar criterion was considered by \cite{Tang-Leng.10}  for a model without missing data, for $\tau=1/2$  and still  with the SCAD penalty.
\section{Algorithms for calculating the estimators}
\label{sect_alg}
In this section we present algorithms to implement the estimation methods in the two previous sections. More precisely, in Subsection \ref{subsect_alg_nonpenal} we present two algorithms  for calculating the smoothed expectile MEL estimator which is solution of problem  (\ref{bnr}) based on the proof of Theorem  \ref{Lemma 2.1 OA} and the adaptive LASSO smoothed expectile MEL estimator solution of (\ref{benr}). For all these algorithms we need a starting point $\eb^{(0)}$. Then the following algorithms, starts with the expectile estimator $\eb^{(0)}=\widetilde{\eb}_n$ given by relation (\ref{tbi}). The algorithms will stop when $\|  \eb^{(k+1)} -\eb^{(k)}\| < \nu$, with $\nu$ fixed at the beginning. We will denote by $\eb^{(k)}$ the value of $\eb$ and  $\el^{(k)}$ that of $\el$ calculated by an algorithm at step $k$. \\
For the four algorithms that will follow we will only give the formula for calculating $\eb^{(k+1)}$ and $\el^{(k+1)}$ knowing their values of the previous steps.
\subsection{Algorithms for smoothed expectile MEL estimator}
\label{subsect_alg_nonpenal}
Here, we propose Algorithms \textit{A1} and \textit{A2} in order to calculate the estimator $\widehat \eb_n$.\\
First of all, let us recall the definitions of $\textbf{S}_1(\eb,\el(\eb))$ and of  $\textbf{S}_2(\eb,\el(\eb))$ given  in Theorem \ref{Lemma 2.1 OA}. From these we calculate their partial derivatives 
\begin{equation*}
\left\{
\begin{split}
\frac{\partial \textbf{S}_{1}	(\eb, \el(\eb))}{\partial \el} &= -\frac{2}{n} \sum^n_{i=1}\frac{ \widehat \eg_i (\eb) \widehat \eg_i (\eb)^\top}{\big(1+\el^\top \widehat \eg_i (\eb) \big)^2}\\
\frac{\partial \textbf{S}_{1}	(\eb, \el(\eb))}{\partial \eb} &=\frac{2}{n} \sum^n_{i=1} \bigg(\frac{\frac{\partial \widehat \eg_i (\eb)}{\partial \eb}}{ 1+\el^\top \widehat \eg_i (\eb)  }-\frac{ \el^\top \frac{\partial \widehat \eg_i (\eb)}{\partial \eb}\widehat \eg_i (\eb) }{\big(1+\el^\top \widehat \eg_i (\eb) \big)^2} \bigg),
\end{split}
\right.
\end{equation*}	
and
\begin{equation*}
\left\{
\begin{split}
\frac{\partial \textbf{S}_{2}(\eb, \el(\eb))}{\partial \el^\top}&=\frac{2}{n} \sum^n_{i=1} \bigg(\frac{\frac{\partial \widehat \eg_i (\eb)}{\partial \eb}}{ 1+\el^\top \widehat \eg_i (\eb)  }-\frac{ \el^\top \frac{\partial \widehat \eg_i (\eb)}{\partial \eb}\widehat \eg_i (\eb) }{\big(1+\el^\top \widehat \eg_i (\eb) \big)^2} \bigg)= 	\frac{\partial \textbf{S}_{1}	(\eb, \el(\eb))}{\partial \eb},\\
\frac{\partial \textbf{S}_{2}(\eb, \el(\eb))}{\partial \eb^\top} & =\frac{2}{n} \sum^n_{i=1} \el^\top \bigg( \frac{\frac{\partial^2 \widehat \eg_i (\eb)}{\partial \eb^2}}{ 1+\el^\top \widehat \eg_i (\eb)  }-\frac{ \el^\top \frac{\partial \widehat \eg_i (\eb)}{\partial \eb} \big(\frac{\partial \widehat \eg_i (\eb)}{\partial \eb}  \big)^\top }{\big(1+\el^\top \widehat \eg_i (\eb) \big)^2}  \bigg).
\end{split}
\right.
\end{equation*}
Thus, we can approximate $\frac{\partial \textbf{S}_{1}	(\eb, \el(\eb))}{\partial \el}$ by $ -2 n^{-1} \sum^n_{i=1}\widehat \eg_i (\eb) \widehat \eg_i (\eb)^\top $,  $\frac{\partial \textbf{S}_{1}	(\eb, \el(\eb))}{\partial \eb}=\frac{\partial \textbf{S}_{2}(\eb, \el(\eb))}{\partial \el^\top}$ by $ -2 n^{-1} \sum^n_{i=1} \frac{\partial \widehat \eg_i (\eb)}{\partial \eb} \big( 1- \el^\top\widehat \eg_i (\eb) \big)$ and $\frac{\partial \textbf{S}_{2}(\eb, \el(\eb))}{\partial \eb^\top} $ by $ -2 n^{-1} \sum^n_{i=1} \el^\top \bigg(\frac{\partial^2 \widehat \eg_i (\eb)}{\partial \eb^2} - \frac{\partial \widehat \eg_i (\eb)}{\partial \eb} \big(\frac{\partial \widehat \eg_i (\eb)}{\partial \eb}  \big)^\top  \bigg) $.\\
Thus, in order to calculate the estimator $\widehat \eb_n$ of (\ref{bnr}) we propose the following two algorithms. \\
\textbf{Algorithm 1} (\underline{called \textit{A1}}). 
From relation (\ref{ZW1})  we calculate at step $k$:
\[
\el^{(k)}\equiv  \el(\eb^{(k)})= \bigg( \frac{1}{n} \sum^n_{i=1} \widehat{\eg}_i(\eb^{(k)})  \widehat{\eg}_i(\eb^{(k)})^\top  \bigg)^{-1} \bigg( \frac{1}{n} \sum^n_{i=1} \widehat{\eg}_i(\eb^{(k)})\bigg).
\]
In order to calculate $\eb^{(k+1)}$ at step $k+1$, knowing $\eb^{(k)}$ and $\el^{(k)}$, using Theorem \ref{Lemma 2.1 OA} and a Taylor expansion, we have:
	\begin{equation*}
	\left\{
	\begin{split}
\oo_p & \simeq\textbf{S}_1\big(\eb^{(k)}, \oo_p\big) - \big( \eb^{(k+1)} -\eb^{(k)} \big)\frac{\partial \textbf{S}_1(\eb^{(k)},\oo_p)}{\partial \eb}\\
 & = \frac{2}{n}\sum^n_{i=1}\widehat {\eg}_i(\eb^{(k)})  -  \big( \eb^{(k+1)} -\eb^{(k)} \big) \frac{2}{n} \sum^n_{i=1} \frac{\partial \widehat {\eg}_i(\eb^{(k)  })}{\partial \eb} \big(  -1 + \el^{(k)^\top}\widehat {\eg}_i(\eb^{(k)})\big).
			\end{split}
	\right.
\end{equation*}
Then, for this algorithm, we calculate at step $(k+1)$:
\begin{equation}
	\label{A1}
	\eb^{(k+1)}=\eb^{(k)} +\frac{1}{n}\sum^n_{i=1} \widehat {\eg}_i(\eb^{(k)  }) \bigg( \frac{1}{n}\sum^n_{i=1}\frac{\partial \widehat {\eg}_i(\eb^{(k)  })}{\partial \eb}  \big(  -1 + \el^{(k)^\top}\widehat {\eg}_i(\eb^{(k)})\big)  \bigg)^{-1}, 
\end{equation}
with 
\[
\el^{(k+1)}\equiv  \el(\eb^{(k+1)})= \bigg( \frac{1}{n} \sum^n_{i=1} \widehat{\eg}_i(\eb^{(k+1)})  \widehat{\eg}_i(\eb^{(k+1)})^\top  \bigg)^{-1} \bigg( \frac{1}{n} \sum^n_{i=1} \widehat{\eg}_i(\eb^{(k+1)})\bigg).
\]
\textbf{Algorithm 2} (\underline{called \textit{A2}})  of Newton-Raphson (see also \cite{Ren-Zhang.11}). \\
Since by Theorem  \ref{Th1 ZW}, $\widehat \el_n \overset{\eP} {\underset{n \rightarrow \infty}{\longrightarrow}} \oo_p$, we propose for this algorithm   not to calculate $\el^{(k)}$, but to always take it $\oo_p$.\\
The $(k+1)$th update $\eb^{(k+1)}$ for $\eb^{(k)} $, using relation (\ref{pap}), is assigned by:
	\begin{align}
		\label{A2}
\eb^{(k+1)} & =\eb^{(k)}- \bigg( \frac{\partial \textbf{S}_1(\eb^{(k)},\oo_p)}{\partial \eb}\bigg)^{-1}\bigg(\frac{1}{n}\sum^n_{i=1} \widehat {\eg}_i(\eb^{(k)  })  \bigg) \nonumber \\  
& =  \eb^{(k)}- \bigg(\frac{1}{n}\sum^n_{i=1} \frac{\partial \widehat {\eg}_i(\eb^{(k)  }) }{\partial \eb}\bigg)^{-1}\bigg(\frac{1}{n}\sum^n_{i=1} \widehat {\eg}_i(\eb^{(k)  })  \bigg).
	\end{align}
These two algorithms will be compared by simulations in Subsection \ref{subsect_complete}.
\subsection{Algorithms for adaptive LASSO smoothed expectile MEL estimator}
\label{subsect_alg_aLASSO}
The following Algorithms \textit{L1} and \textit{L2} are  based on relations obtained in the proof of Theorem  \ref{thh} and Remark \ref{vit_el}.
Moreover, in the following two algorithms, if $|\beta^{(k+1)}_j| < \epsilon$, with   $\epsilon$ a preset value, then we set $\beta^{(m)}_j =0$ for any $m \geq k+1$.\\
\textbf{Algorithm 1} (\underline{called \textit{L1}}).  The penalty $p(\eb_j)\equiv n \eta_n \widehat \omega_{n,j} |\beta_j|$ is written as in \cite{Fan-Li.01}: $p(||\beta_j|)=p(|\beta^0_j|)+2^{-1}  p'(|\beta^0_j|)/|\beta^0_j| (\beta^2_j - \beta^{0^2}_j)$, with $p'(|\beta^0_j|)=n \eta_n \widehat \omega_{n,j} $. Then, let us consider the diagonal matrix $\textbf{D}(\eb) \equiv \textrm{diag} \big(\frac{\eta_n \widehat \omega_{n,j}}{|\beta_j|}  \big)_{1 \leqslant j \leqslant p}$. Thus, similarly as for algorithm \textit{A1}, we consider:
\[
\oo_p \simeq \textbf{S}_1(\eb^{(k)},\oo_p)+ \big( \eb^{(k+1)} -\eb^{(k)} \big)\frac{\partial \textbf{S}_1(\eb^{(k)},\oo_p)}{\partial \eb} +\textbf{D}(\eb^{(k)}) \eb^{(k+1)},
\]
from where
\begin{equation}
	\label{AL1}
	\begin{split}
	\eb^{(k+1)}=&\eb^{(k)}+ \bigg(\frac{1}{n}\sum^n_{i=1}  \frac{\partial \widehat {\eg}_i(\eb^{(k)  })  }{\partial \eb}\big(\el^{(k)^\top}\widehat {\eg}_i(\eb^{(k)  }) -1\big)- \textbf{D}(\eb^{(k)})\bigg)^{-1} \\
	 & \qquad \qquad \qquad \cdot \bigg(\frac{1}{n}\sum^n_{i=1} \widehat {\eg}_i(\eb^{(k)  })+\textbf{D}(\eb^{(k)}) \eb^{(k)} \bigg).
	\end{split}
\end{equation}
Since $\widehat \el^*_n$ is solution like $\widehat \el_n$ of equation (\ref{LL}), then we consider the following  tuning parameter $\el^{(k+1)}$ at step $(k+1)$:
\[
\el^{(k+1)}\equiv  \el(\eb^{(k+1)})= \bigg( \frac{1}{n} \sum^n_{i=1} \widehat{\eg}_i(\eb^{(k+1)})  \widehat{\eg}_i(\eb^{(k+1)})^\top  \bigg)^{-1} \bigg( \frac{1}{n} \sum^n_{i=1} \widehat{\eg}_i(\eb^{(k+1)})\bigg).
\]
\textbf{Algorithm 2} (\underline{called \textit{L2}}).  For this algorithm the tuning parameter is  $\el^{(k)}=\oo_k$ for all $k$ and $	\eb^{(k+1)} $ calculated by
\begin{equation}
\label{AL1bis}
\eb^{(k+1)}=\eb^{(k)}- \bigg(\frac{1}{n}\sum^n_{i=1}  \frac{\partial \widehat {\eg}_i(\eb^{(k)  })  }{\partial \eb}+ \textbf{D}(\eb^{(k)})\bigg)^{-1} \bigg(\frac{1}{n}\sum^n_{i=1} \widehat {\eg}_i(\eb^{(k)  })  +\textbf{D}(\eb^{(k)}) \eb^{(k)} \bigg).
\end{equation}
For algorithms \textit{L1} and \textit{L2}, to each step  we take  $\widehat \omega_{n,j}=| \widehat \beta_{n,j}|^{-\gamma}$, with $\widehat \eb_{n}$ the smoothed expectile MEL estimation calculated on an independent data set by algorithms \textit{A1} and \textit{A2}, respectively. Algorithms \textit{L1} and \textit{L2} will be compared by simulations in Subsection \ref{subsect_complete}. \\

\noindent And finally, note that for all these four algorithms we need  $\partial \widehat {\eg}_i(\eb) / \partial \eb$. By elementary calculus, we obtain that, for any $i=1, \cdots, n$:
 	\begin{align*}
 \frac{\partial  \widehat {\eg}_i(\eb) }{ \partial \eb}&=\delta_i \bigg( \frac{1-2 \tau}{h}K\bigg(\frac{\eX_i^\top \eb - Y_i}{h}\bigg) (Y_i - \eX_i^\top \eb) - \psi_h(\eX_i,Y_i; \eb)\bigg) \eX_i \eX_i^\top \\
 & = \delta_i \bigg( \frac{1-2 \tau}{h}K\bigg(\frac{\eX_i^\top \eb - Y_i}{h}\bigg) (Y_i - \eX_i^\top \eb) -  \big( \tau+(1-2 \tau) \int_{v< \frac{\eX_i^\top \eb -Y_i}{h}}K(v)dv\big)\bigg) \eX_i \eX_i^\top .
\end{align*}
\section{Simulation studies}
\label{section_simu}
In this section we perform a numerical simulation study to illustrate the theoretical results which are put into practice via the two proposed algorithms to find the smoothed  expectile MLE estimations and another two algorithms to calculate the adaptive LASSO smoothed expectile MEL estimations. This study is also used to compare the four algorithms proposed in Section \ref{sect_alg}. All simulations were performed under the R language. Note that in order to calculate the expectile estimation necessary as a starting point for each of the algorithms, we use the  function \textit{ernet} of  the R package \textit{SALES}.\\
The true values of the parameters are $\ebo=(\beta^0_1, \cdots , \beta^0_p)$ with $\beta^0_3=1$, $\beta^0_5=2$ and   in some cases $\beta^0_7=-1$,  the other elements being equal with 0.  We consider  $\gamma=2.5$, $h=n^{-1/4}$, $\eta_n=n^{-5/6}$, $\nu=10^{-2}$, $\alpha=0.05$ and for the two algorithms of Subsection \ref{subsect_alg_aLASSO} we take $\epsilon=10^{-4}$. The values of $n$ are in $\{100, 200, 300, \cdots, 2000\}$. 
Two designs $(X_{ji})_{1 \leqslant j \leqslant p; 1 \leqslant i \leqslant n}$ are considered:
\begin{description}
	\item  ${\cal D}1$: $X_{ji} \sim {\cal N}(0,1)$, for any $j\in \{1, \cdots, p\}$, $i \in \{1, \cdots n\}$;
	\item   ${\cal D}2$: $X_{ji} \sim \chi^2(1)+j^2/n$, for any $j\in \{1, \cdots, p\}\setminus \{3\}$, $i \in \{1, \cdots n\}$.
\end{description}
 Two distributions are considered for model errors: standard normal distribution denoted ${\cal N}(0,1)$ and exponential distribution with mean -1.5 denoted  ${\cal E}\!xp(-1.5)$.
We calculate an approximation of the expectile empirical log-likelihood ratio $\widehat{\cal R}_n(\ebo, \el(\ebo))$ using relation (\ref{ZW2}):
\begin{equation}
	\label{ZW2_aprox}
	\widehat{\cal R}_n(\ebo, \el(\ebo))  \simeq n \bigg(\frac{1}{n} \sum^n_{i=1} \widehat{\eg}_i(\ebo)^\top\bigg) \bigg( \frac{1}{n} \sum^n_{i=1} \widehat{\eg}_i(\ebo)  \widehat{\eg}_i(\ebo)^\top \bigg)^{-1} \bigg(\frac{1}{n} \sum^n_{i=1} \widehat{\eg}_i(\ebo)\bigg)
\end{equation}
and of the process $\widehat{\cal R}_n(\eb, \el(\eb))$ using (\ref{LUb}):
\begin{equation} 
	\label{LUb_aprox}
	\widehat{\cal R}_n(\eb, \el(\eb))  \simeq n \bigg(\frac{1}{n} \sum^n_{i=1} \widehat{\eg}_i(\eb)^\top\bigg) \bigg( \frac{1}{n} \sum^n_{i=1} \widehat{\eg}_i(\eb)  \widehat{\eg}_i(\eb)^\top \bigg)^{-1} \bigg(\frac{1}{n} \sum^n_{i=1} \widehat{\eg}_i(\eb)\bigg).
\end{equation}
The confidence region (CR) of parameters is calculated combining Theorem \ref{Th2 ZW}  with the two unpenalized estimations obtained by algorithms \textit{A1} and \textit{A2} of Subsection \ref{subsect_alg_nonpenal}. If in addition we want to perform automatic selection of non-zero coefficients, then the CR is calculated by combining  two penalized estimations obtained by algorithms \textit{L1} and \textit{L2} of Subsection \ref{subsect_alg_aLASSO} with Corollary \ref{corrolaire_RnA}, for which we take the quantile of $\chi^2(|\widehat{\cal A}^*_n|)$ distribution.
Therefore, the confidence region (CR) for an estimation method is:
\begin{equation}
	\label{CR22}
CR \equiv \big\{\eb ; \;\; \widehat{\cal R}_n(  \eb, \el( \eb)) \leq c_{1-\alpha;dl} \big\},
\end{equation}
where $\widehat{\cal R}_n(\eb, \el(\eb))$ is calculated by (\ref{LUb_aprox}), $dl$ is either $p$ for smoothed expectile MEL estimations and for CP or $|\widehat {\cal A}^*_n|$ for adaptive LASSO smoothed expectile MEL estimations, while  $c_{1-\alpha;dl}$ is the $(1 -\alpha)$ quantile of $\chi^2(dl)$ distribution. Note that for each CR  we calculate by M Monte Carlo replications:
 \begin{equation}
	\label{alf}
	1- \widehat \alpha \equiv \frac{1}{M} \sum^M_{i=1} \e1_{\widehat{\cal R}_n(\widehat \eb_n, \el(\widehat \eb_n)) \leq c_{1-\alpha;dl}},
\end{equation}
where  $\widehat \eb_n$ is the estimation of $\eb$ obtained by one of the four algorithms. When $1- \widehat{\alpha}$ is calculated for $\widehat{\cal R}_n(\ebo, \el(\ebo))  $ we obtain the coverage probability (CP) that is the frequency with which  the true value $\ebo$ tails into the confidence region by Monte Carlo replications.  We will therefore build the CR on the basis of the estimations obtained by algorithms which in turn correspond to   estimation methods of $\eb$. For the comparison purpose, we also calculate $\widehat{\cal R}_n(\ebo, \el(\ebo))$ using  relation (\ref{ZW2_aprox}), in which case we denote the CR by CR0.  \\
The study  results are presented in Subsections \ref{subsect_complete} and \ref{subsect_missing} and the conclusion of simulations in Subsection \ref{subsect_conclusion}.
\subsection{Complete data}
\label{subsect_complete}
In this subsection the data of the response variable are considered complete. 
\begin{table}
	\caption{\footnotesize Results  for $\| \widehat \eb_n -\ebo \| $, $1-  \widehat \alpha$, $   |{\widehat {\cal A}}^{*c}_n|/|{\cal A}^c| $ obtained by 1000 Monte Carlo replications when the data are complete, by four algorithms, when $\ebo=(\beta^0_1, \cdots , \beta^0_p)$ with $\beta^0_3=2$, $\beta^0_5=1$, for $p \in \{5, 10\}$.}
	\begin{center}
		{\scriptsize
			\begin{tabular}{|c|c|c|c|cccc|ccccc|cc|}\hline  
			$\varepsilon$ & $\eX_i$ & $p$  &	$n$	&  \multicolumn{4}{c|}{$\| \widehat \eb_n -\ebo \| $}  & \multicolumn{5}{c|}{$1-  \widehat \alpha$} & \multicolumn{2}{c|}{$   |{\widehat {\cal A}}^{*c}_n|/|{\cal A}^c| $} \\ 
				\cline{5-15} 
				&  &   &   &  \textit{A1} & \textit{A2} & \textit{L1} & \textit{L2} &\textit{CP} &  \textit{A1} & \textit{A2} & \textit{L1} & \textit{L2}  & \textit{L1} & \textit{L2}    \\ \hline \hline
				 ${\cal E}\!xp(-1.5)$ & ${\cal D}1$   & 5 & 100 &  0.28  & 0.28    & 0.13   & 0.13   &  0.89  & 1  & 1  & 0.95  & 0.95   &  0.91 & 0.91   \\ 
				 &  &   &  500 &  0.13  &  0.13   & 0.06  & 0.06  &  0.95  &  1 & 1  & 0.91  &  0.91  &  0.97 &  0.97  \\ 
				  &  &   &  1000 &  0.09  &  0.09   & 0.04  & 0.04  &  0.95  &  1 & 1  & 0.93  &  0.93  &  0.99 &  0.99  \\ \cline{3-15} 
				   &     & 10 & 100 &  0.43  & 0.43   & 0.14   & 0.14   &  0.76  & 1  & 1  & 0.79  & 0.79   &  0.93 & 0.93   \\ 
				  &  &   &  500 &  0.20  &  0.20   & 0.06  & 0.06  &  0.91  &  1 & 1  & 0.66  &  0.66  &  0.97 &  0.97  \\ 
				  &  &   &  1000 &  0.14  &  0.14   & 0.04  & 0.04  &  0.93  &  1 & 1  & 0.74  &  0.74  &  0.99 &  0.99  \\ \cline{2-15} 
				  & ${\cal D}2$   & 5 & 100 &  0.22  & 0.22    & 0.10   & 0.10   &  0.86  & 1  & 1  & 0.89  & 0.89   &  0.94 & 0.94   \\ 
				  &  &   &  500 &  0.10  &  0.10   & 0.04  & 0.04  &  0.93  &  1 & 1  & 0.91  &  0.91  &  0.99 &  0.99  \\ 
				  &  &   &  1000 &  0.07  &  0.07   & 0.03  & 0.03  &  0.95  &  1 & 1  & 0.95  &  0.95  &  0.99 &  0.99  \\ \cline{3-15} 
				  &     & 10 & 100 &  0.33  & 0.33   & 0.10   & 0.10   &  0.72  & 1  & 1  & 0.75  & 0.75  &  0.96 & 0.96  \\ 
				  &  &   &  500 &  0.15  &  0.15   & 0.04  & 0.04  &  0.85  &  1 & 1  & 0.69  &  0.69  &  0.99 &  0.99  \\ 
				  &  &   &  1000 &  0.10  &  0.10   & 0.03  & 0.03  &  0.91  &  1 & 1  & 0.81  &  0.81  &  0.99 &  0.99  \\ \cline{1-15}
				  ${\cal N}(0,1)$ & ${\cal D}1$   & 5 & 100 &  0.19  & 0.19    & 0.09   & 0.09   &  0.99  & 1  & 1  & 0.93  & 0.93   &  0.95 & 0.95   \\ 
				  &  &   &  500 &  0.08  &  0.08   & 0.04  & 0.04  &  0.98  &  1 & 1  & 0.96  &  0.96  &  0.99 & 0.99  \\ 
				  &  &   &  1000 &  0.06  &  0.06   & 0.03  & 0.03  &  0.97  &  1 & 1  & 0.97  &  0.97  &  1 &  1  \\ \cline{3-15} 
				  &     & 10 & 100 &  0.31  & 0.31   & 0.09   & 0.09  &  0.98  & 1  & 1  & 0.62  & 0.62   &  0.95 & 0.95   \\ 
				  &  &   &  500 &  0.13  &  0.13   & 0.04  & 0.04  &  0.98  &  1 & 1  & 0.81  &  0.81  &  0.99 &  0.99  \\ 
				  &  &   &  1000 &  0.09  &  0.09   & 0.02  & 0.02  &  0.96  &  1 & 1  & 0.90  &  0.90  &  0.99 & 0.99  \\ \cline{2-15} 
				  & ${\cal D}2$   & 5 & 100 &  0.14  & 0.14  & 0.07   & 0.07   &  0.98  & 1  & 1  & 0.95  & 0.95   &  0.98 & 0.98  \\ 
				  &  &   &  500 &  0.06  &  0.06   & 0.03  & 0.03  &  0.97  &  1 & 1  & 0.98  &  0.98  &  1 &  1  \\ 
				  &  &   &  1000 &  0.04  &  0.04   & 0.02  & 0.02  &  0.97  &  1 & 1  & 0.99  &  0.99  &  1 & 1  \\ \cline{3-15} 
				  &     & 10 & 100 &  0.24  & 0.24   & 0.07   & 0.07   &  0.98  & 1  & 1  & 0.84  & 0.84   &  0.97 & 0.97   \\ 
				  &  &   &  500 &  0.09  &  0.09  & 0.03  & 0.03  &  0.97  &  1 & 1  & 0.90  &  0.90  &  1&  1  \\ 
				  &  &   &  1000 &  0.06  &  0.06   & 0.02  & 0.02  &  0.97  &  1 & 1  & 0.95  &  0.95  &  1&  1  \\ \cline{1-15}
				 		\end{tabular} 
		}
	\end{center}
\label{Tab1} 
\end{table}
For  $\beta^0_3=1$, $\beta^0_5=2$ and either $\beta^0_j=0$ or $\beta^0_j=1$ for  the others $j$,  in Tables \ref{Tab1} and \ref{Tab2}, respectively, for $M=1000$ Monte Carlo replications, we present  $1-\widehat \alpha $ calculated by (\ref{alf}) using (\ref{LUb_aprox}), the norm $\| \widehat \eb_n -\ebo \| $ for the precision of the parameters estimations obtained by the four algorithms and an indicator$|{\widehat {\cal A}}^{*c}_n|/|{\cal A}^c| $ for the automatic detection of null parameters by the  two algorithms \textit{L1} and \textit{L2}.
From Table \ref{Tab1} we deduce that there is no difference between  algorithms \textit{A1} and \textit{A2}. Moreover, there is also no difference between the obtained results by  \textit{L1} and \textit{L2}. Consequently, we give up algorithms \textit{A1} and  \textit{L1}, keeping those that are calculated simpler: \textit{A2} and \textit{L2}.  There are no significant differences in the results  obtained for the two designs ${\cal D}1$ and ${\cal D}2$ considered. On the other hand, the results are better when $\varepsilon \sim {\cal N}(0,1)$ even if they remain very good also for $\varepsilon \sim {\cal E}\!xp(-1.5)$ but for larger $n$. For algorithm \textit{L2}, it is necessary first to eliminate the variables with the zero estimated coefficients and then to start again with the \textit{A2} algorithm, whereupon  $1-  \widehat \alpha$ is close to 1 when all the coefficients are non-zero. This idea is supported by the results obtained in Table  \ref{Tab2}. More precisely, in  Table \ref{Tab2}, where all true coefficients are non zero,  we observe that by    algorithm \textit{L2} we don't obtain any null estimation (false null) since $   |{\widehat {\cal A}}^{*c}_n|/|{\cal A}^c| $ is always NaN.
\begin{table}
	\caption{\footnotesize Results    for $\| \widehat \eb_n -\ebo \| $, $1-  \widehat \alpha$, $   |{\widehat {\cal A}}^{*c}_n|/|{\cal A}^c| $ obtained by 1000 Monte Carlo replications when the data are complete,  by two algorithms \textit{A2} and \textit{L2}, when $\ebo=(\beta^0_1, \cdots , \beta^0_p)$ with $\beta^0_3=2$, $\beta^0_5=1$, and $\beta_j=1$ for the other $j$. The model errors are $\varepsilon \sim {\cal E}\!xp(-1.5)$.}
	\begin{center}
		{\scriptsize
			\begin{tabular}{|c|c|c|cc|ccc|c|}\hline  
				 $\eX_i$ & $p$  &	$n$	&  \multicolumn{2}{c|}{$\| \widehat \eb_n -\ebo \| $}  & \multicolumn{3}{c|}{$1-  \widehat \alpha$} & \multicolumn{1}{c|}{$   |{\widehat {\cal A}}^{*c}_n|/|{\cal A}^c| $} \\ 
				\cline{4-9} 
				  &   &   &    \textit{A2}   & \textit{L2} &\textit{CP} &    \textit{A2} &   \textit{L2}  &  \textit{L2}    \\ \hline \hline
			 ${\cal D}1$   & 5 & 100 &  0.28  & 0.32  &  0.89  & 1  & 0.99    & NaN   \\ 
			 &  & 500 &  0.13  & 0.13  &  0.94  & 1  & 1    & NaN   \\ \cline{2-9}
			   & 10 & 100 &  0.44  & 0.51  &  0.75  & 1  & 0.98    & NaN   \\ 
			 &  & 500 &  0.20  & 0.21  &  0.90  & 1  & 1    & NaN   \\ \hline
						  ${\cal D}2$    & 5 & 100 &  0.22 & 0.22  &  0.85  & 1  & 0.99    & NaN   \\ 
						  &  & 500 &  0.09  & 0.09  &  0.93  & 1  & 1    & NaN   \\ \cline{2-9}						  
						   & 10 & 100 &  0.33  & 0.34  &  0.72  & 1  & 1    & NaN   \\ 
			 &  & 500 &  0.15  & 0.15  &  0.86  & 1  & 1    & NaN   \\ \hline
					\end{tabular} 
			}
		\end{center}
	\label{Tab2} 
\end{table}
 
\begin{figure*} 
\begin{minipage}[b]{0.45\linewidth}
	\centering \includegraphics[scale=0.35]{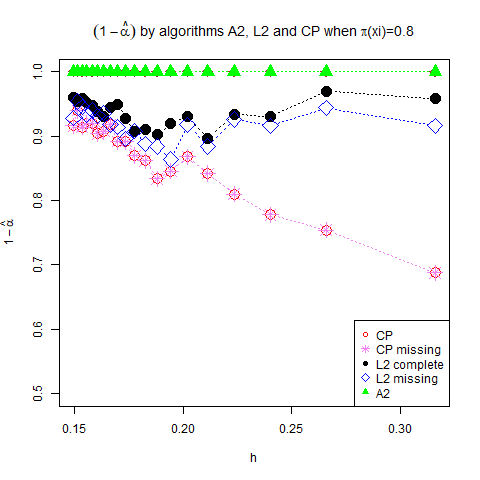}
	\caption{\it   $(1-\widehat \alpha)$ versus $h$ for CRs calculated by (\ref{LUb_aprox}) and (\ref{CR22}),  using  A2,  L2 and CP  for CR0 when $Y$ missing (for $\pi(\ex_i)=0.8$) or complete. $p=10$, $\varepsilon \sim {\cal E}\!xp(-1.5)$, $X \sim {\cal D}2$, ${\cal A}=\{1,2,7\}$.}
	\label{fig_CR_h_miss}
\end{minipage} \hfill
\begin{minipage}[b]{0.45\linewidth}
	\centering \includegraphics[scale=0.35]{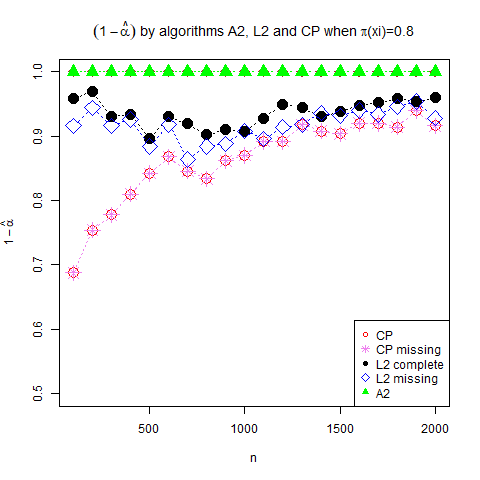}
	\caption{ \it $(1-\widehat \alpha)$ versus $n$ for CRs calculated by (\ref{LUb_aprox}) and (\ref{CR22}),  using   A2,  L2 and CP  for CR0 when  $Y$ missing (for $\pi(\ex_i)=0.8$) or complete. $p=10$, $\varepsilon \sim {\cal E}\!xp(-1.5)$, $X \sim {\cal D}2$, ${\cal A}=\{1,2,7\}$.}
	\label{fig_CR_n_miss}  
\end{minipage} \hfill
		\begin{minipage}[b]{0.45\linewidth}
		\centering \includegraphics[scale=0.35]{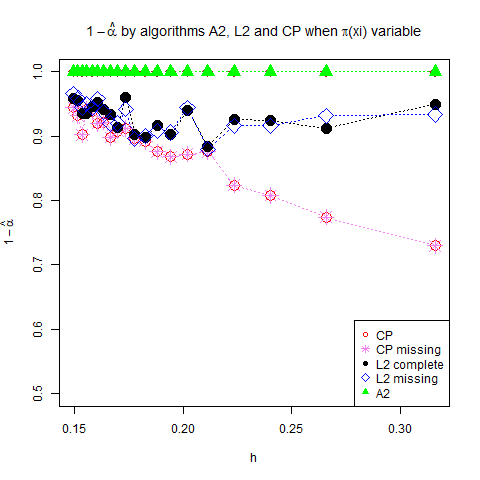}
		\caption{\it  $(1-\widehat \alpha)$ versus $h$ for CRs calculated by (\ref{LUb_aprox}) and (\ref{CR22}),  using   A2,  L2 and CP  for CR0 when  $Y$ missing (for $\pi(\ex_i)$ variable with $\ex_i$) or complete. $p=10$, $\varepsilon \sim {\cal E}\!xp(-1.5)$, $X \sim {\cal D}2$, ${\cal A}=\{1,2,7\}$.}
		\label{fig_CR_h_miss_pivar}
	\end{minipage} \hfill
	\begin{minipage}[b]{0.45\linewidth}
		\centering \includegraphics[scale=0.35]{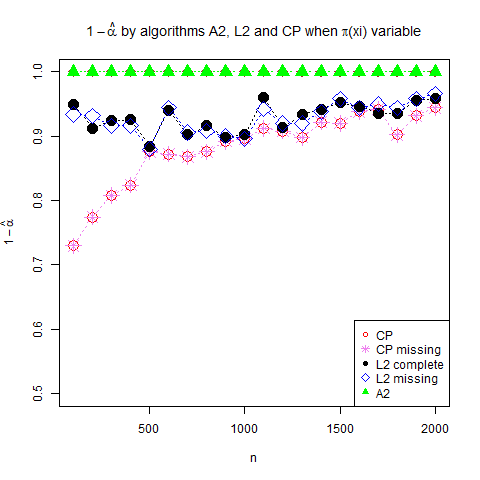}
		\caption{ \it $(1-\widehat \alpha)$ versus $n$ for CRs calculated by (\ref{LUb_aprox})  and (\ref{CR22}),  using   A2,  L2 and CP  for CR0 when $Y$ missing (for $\pi(\ex_i)$ variable with $\ex_i$) or complete. $p=10$, $\varepsilon \sim {\cal E}\!xp(-1.5)$, $X \sim {\cal D}2$, ${\cal A}=\{1,2,7\}$.}
		\label{fig_CR_n_miss_pivar}  
	\end{minipage}
\end{figure*}

\begin{figure*} 
\begin{minipage}[b]{0.45\linewidth}
	\centering \includegraphics[scale=0.35]{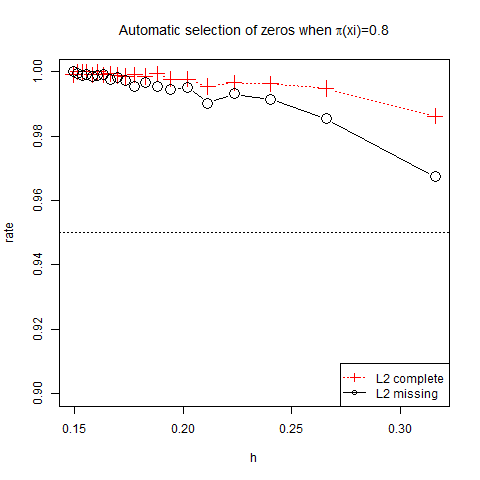}
	\caption{\it  Automatic selection rate of non-zero coefficients versus $h$ when  $Y$ missing (for $\pi(\ex_i)=0.8$) or complete. $p=10$, $\varepsilon \sim {\cal E}\!xp(-1.5)$, $X \sim {\cal D}2$, ${\cal A}=\{1,2,7\}$.}
	\label{fig_AnA_h_miss}
\end{minipage} \hfill
\begin{minipage}[b]{0.45\linewidth}
	\centering \includegraphics[scale=0.35]{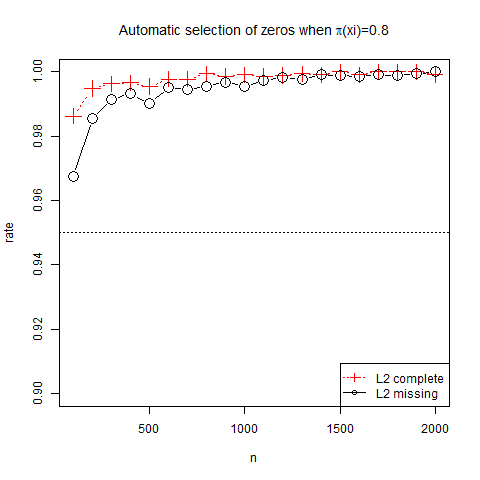}
	\caption{ \it Automatic selection rate of non-zero coefficients versus $n$ when  $Y$ missing (for $\pi(\ex_i)=0.8$) or complete. $p=10$, $\varepsilon \sim {\cal E}\!xp(-1.5)$, $X \sim {\cal D}2$, ${\cal A}=\{1,2,7\}$. }
	\label{fig_AnA_n_miss}  
\end{minipage}  \hfill
	\begin{minipage}[b]{0.45\linewidth}
	\centering \includegraphics[scale=0.35]{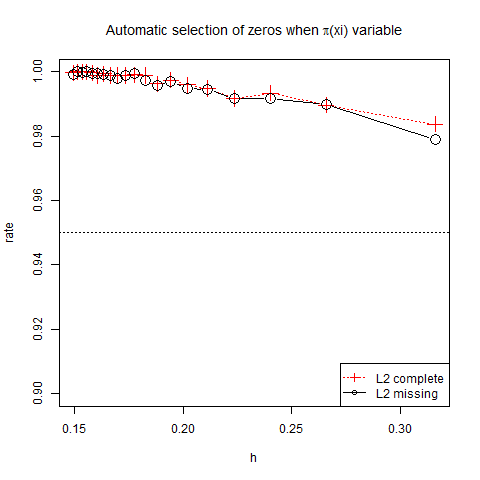}
	\caption{\it  Evolution with respect to $h$  of the automatic selection rate of non-zero coefficients when $Y$ missing (for $\pi(\ex_i)$ variable with $\ex_i$) or complete. $p=10$, $\varepsilon \sim {\cal E}\!xp(-1.5)$, $X \sim {\cal D}2$, ${\cal A}=\{1,2,7\}$.}
	\label{fig_AnA_h_miss_pivar}
\end{minipage} \hfill
\begin{minipage}[b]{0.45\linewidth}
	\centering \includegraphics[scale=0.35]{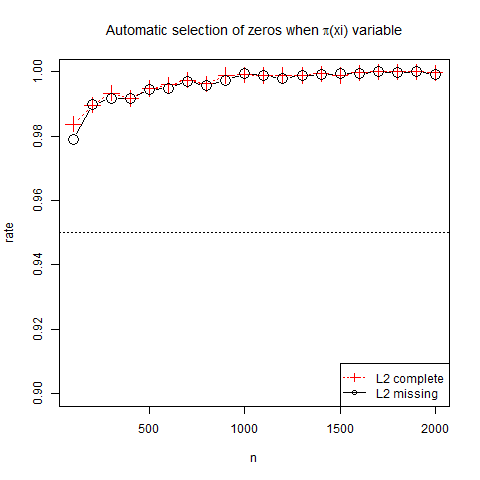}
	\caption{ \it Evolution with respect to $n$ of the automatic selection rate of non-zero coefficients when $Y$ missing (for $\pi(\ex_i)$ variable with $\ex_i$) or complete. $p=10$, $\varepsilon \sim {\cal E}\!xp(-1.5)$, $X \sim {\cal D}2$, ${\cal A}=\{1,2,7\}$.}
	\label{fig_AnA_n_miss_pivar}  
\end{minipage} \hfill
\end{figure*}
\begin{figure*} 
	\begin{minipage}[b]{0.45\linewidth}
		\centering \includegraphics[scale=0.35]{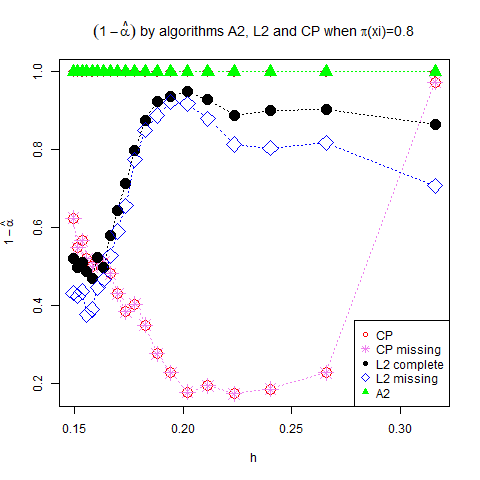}
		\caption{\it   $(1-\widehat \alpha)$ versus $h$ for CRs calculated by (\ref{LUb_aprox})  and (\ref{CR22}),  using  A2,  L2 and CP  for CR0 when $Y$ missing (for $\pi(\ex_i)=0.8$) or complete. $p=50$, $\varepsilon \sim {\cal E}\!xp(-1.5)$, $X \sim {\cal D}2$, ${\cal A}=\{1,2,7\}$.}
		\label{fig_CR_h_miss_p50}
	\end{minipage} \hfill
	\begin{minipage}[b]{0.45\linewidth}
		\centering \includegraphics[scale=0.35]{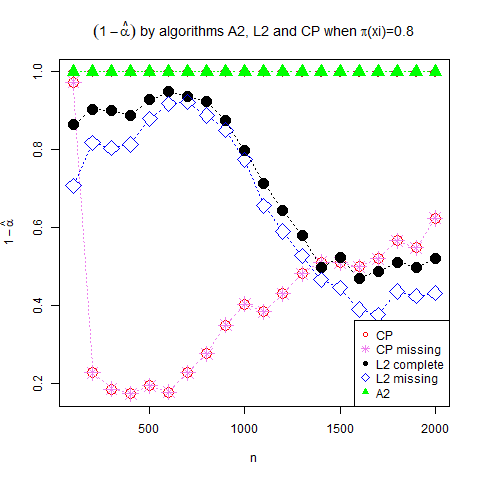}
		\caption{ \it $(1-\widehat \alpha)$ versus $n$ for CRs calculated by (\ref{LUb_aprox})  and (\ref{CR22}),  using   A2,  L2 and CP  for CR0 when  $Y$ missing (for $\pi(\ex_i)=0.8$) or complete. $p=50$, $\varepsilon \sim {\cal E}\!xp(-1.5)$, $X \sim {\cal D}2$, ${\cal A}=\{1,2,7\}$.}
		\label{fig_CR_n_miss_p50}  
	\end{minipage} \hfill
	\begin{minipage}[b]{0.45\linewidth}
		\centering \includegraphics[scale=0.35]{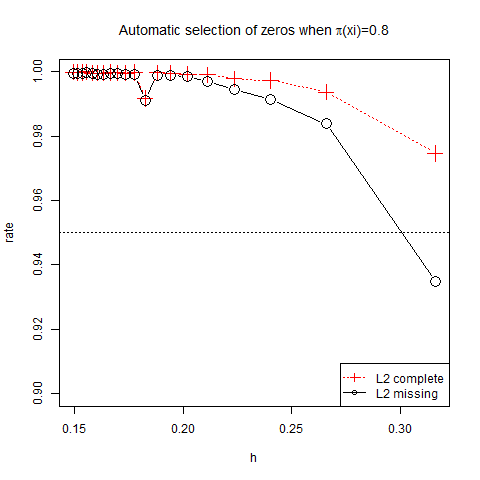}
		\caption{\it  Automatic selection rate of non-zero coefficients versus $h$ when  $Y$ missing (for $\pi(\ex_i)=0.8$) or complete. $p=50$, $\varepsilon \sim {\cal E}\!xp(-1.5)$, $X \sim {\cal D}2$, ${\cal A}=\{1,2,7\}$.}
		\label{fig_AnA_h_miss_p50}
	\end{minipage} \hfill
	\begin{minipage}[b]{0.45\linewidth}
		\centering \includegraphics[scale=0.35]{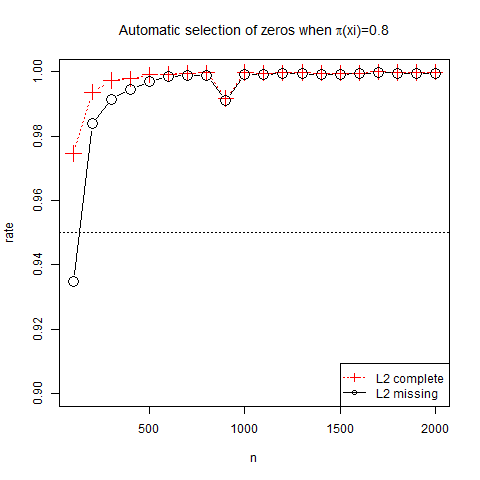}
		\caption{ \it Automatic selection rate of non-zero coefficients versus $n$ when  $Y$ missing (for $\pi(\ex_i)=0.8$) or complete. $p=50$, $\varepsilon \sim {\cal E}\!xp(-1.5)$, $X \sim {\cal D}2$, ${\cal A}=\{1,2,7\}$. }
		\label{fig_AnA_n_miss_p50}  
	\end{minipage}  \hfill
\end{figure*}
\noindent  Figures \ref{fig_CR_h_miss} of \ref{fig_AnA_n_miss_pivar}  were made following 500 Monte Carlo replications  for $p=10$, with $\beta^0_3=1$,  $\beta^0_5=2$,  $\beta^0_7=-1$ and the others $\beta^0_j=0$. The model errors are  ${\cal E}\!xp(-1.5)$ and the design is ${\cal D}2$.
In Figures \ref{fig_CR_h_miss} and \ref{fig_CR_h_miss_pivar} we have, among others, the evolution with respect $h$ of the $(1-\widehat{\alpha})$ corresponding to CRs calculated by (\ref{LUb_aprox})  using algorithms \textit{A2, L2} and the  CP corresponding to CR0.  In Figures \ref{fig_CR_n_miss} and \ref{fig_CR_n_miss_pivar}, the same evolution is presented with respect $n$. In general, $1 - \widehat \alpha \geq 0.9$ by \textit{A2} and \textit{L2} (except for $n=500$) and the value of $1 - \widehat \alpha$ changes very little with $n$. Moreover if the CR is calculated for $\ebo$, the results are less good when $n \leq 1000$ (and then $h$ bigger).  We obtain that   $(1-\widehat{\alpha})$ by algorithm \textit{A2},  is  always equal to 1. In Figures \ref{fig_AnA_h_miss} and \ref{fig_AnA_h_miss_pivar} we find the evolution with respect $h$ of the 	automatic selection rate of non-zero coefficients using algorithm \textit{L2} and in Figures \ref{fig_AnA_n_miss} and \ref{fig_AnA_n_miss_pivar} we observe the evolution in respect to $n$. The three non-zero coefficients are identified at least at $98\%$ for all $n \geq 100$ by algorithm \textit{L2}.  
In Figures \ref{fig_CR_h_miss_p50} of \ref{fig_AnA_n_miss_p50} we present the same indicators as in Figures  \ref{fig_CR_h_miss}, \ref{fig_CR_n_miss}, \ref{fig_AnA_h_miss} and \ref{fig_AnA_n_miss} in the same configuration, except that the number of zeros is larger, more precisely $p-q=|{\cal A}^c|=47$. Note that $1 - \widehat \alpha$ decreases from $n \geq 1000$ by algorithm \textit{L2} but by \textit{A2} it increases. Hence the idea, already obtained, that we must first eliminate the variables with the zero estimated coefficients using \textit{L2} and afterwards apply \textit{A2} to the model with the remaining variables.
\subsection{Missing data}
\label{subsect_missing}
In this subsection, the response variable $Y$ can be missing, with the missing probabilities  $\pi(\ex_i)$ depending or not on the values of $\ex_i$. For  the probability $\pi(\ex_i)=0.8$, in Figures \ref{fig_CR_h_miss} and \ref{fig_CR_n_miss} we present the   behavior of $(1- \widehat{\alpha})$ with respect to $h$ and $n$ for CR0, \textit{A2} and \textit{L2} with missing data for response variable. In Figures \ref{fig_AnA_h_miss} and \ref{fig_AnA_n_miss} we present the evolution of the  automatic selection rate of non-zero coefficients with respect $h$, $n$, when $Y$ has  missing value.
   The same indicators are represented  in Figures \ref{fig_CR_h_miss_pivar}, \ref{fig_CR_n_miss_pivar}, \ref{fig_AnA_h_miss_pivar} and \ref{fig_AnA_n_miss_pivar}  when  the probability $\pi(x_{ji})$ are  $\pi(x_{ji})=(0.8+0.2 |x_{ji}-1|) \e1_{|x_{ji}-1| \leq 1}+0.95  \e1_{|x_{ji}-1| > 1}$, with $j\in \{1, \cdots, p\}$ and $i=1, \cdots , n$.\\
The CPs on complete or missing data are equal, no matter if $\pi(\ex_i)$ depends  or not on $\ex_i$ (see Figures \ref{fig_CR_h_miss}-\ref{fig_CR_n_miss_pivar}).  Concerning the values of $(1-\widehat{\alpha})$ calculated by \textit{L2}, they are very close on the complete data and on with missing data when $h <0.32$ (or $n \geq 100$). Instead, their values  remain approximately constant with respect to $n$ (see  also Figures \ref{fig_CR_h_miss} of \ref{fig_CR_n_miss_pivar}). From Figures \ref{fig_AnA_h_miss} and \ref{fig_AnA_n_miss} we deduce, this is quite logical, that if $20\%$ of the values of $Y$ are missing, then the automatic selection of zero coefficients is less good when the number  $n$ of observations  is low. Contrariwise, if $Y$ is missing because of the values of $\ex_i$, the automatic selection on incomplete data is close to that obtained on missing data, which seems logical because for 2000 observations, only $8\%$ are missing (see Figures \ref{fig_AnA_h_miss} of \ref{fig_AnA_n_miss_pivar}). The detection rate  of true zeros decreases slowly if $h$ increases (or $n$ decreases), but remains greater than $95\%$.
   
   \subsection{Conclusion of the simulations}
   \label{subsect_conclusion}
   If we know that a priori  the linear model does not have non-zero coefficients, then it is appropriate to use algorithm \textit{A2}, otherwise, we must first select the non-zero coefficients by algorithm \textit{L2}, because for $h < 0.27$ the   automatic selection rate of zero coefficients exceeding $95\%$. Afterwards, algorithm \textit{A2} must be applied for the model that will contain the explanatory variables that had non-zero estimates. If the number of explanatory variables is five then the results obtained by  algorithms \textit{A2} and \textit{L2} are similar when $\varepsilon \sim {\cal N}(0,1)$ or $\varepsilon \sim {\cal E}\!xp(-1.5)$, while for $p=10$, the results for $(1-\widehat{\alpha})$  by \textit{L2} are worse when $\varepsilon \sim {\cal E}\!xp(-1.5)$ than when $\varepsilon \sim {\cal N}(0,1)$ especially when $n$ is small.\\
   Comparing Figures \ref{fig_CR_h_miss} and \ref{fig_CR_h_miss_p50} we observe that when $p$ is large ($p=50$) then for $h<0.20$ (or $n >800$), we obtain $1- \widehat \alpha <0.90$, i.e. in less than $90\%$ of cases $\eb$ is found in CR. Another important result deduced from Figures \ref{fig_AnA_h_miss_p50} and \ref{fig_AnA_n_miss_p50} is that even if the number of zero coefficients is large, their automatic selection remains very efficient. From all  figures we also deduce that the result of Theorem \ref{Th2 ZW} is validated by simulations.\\
   If all model coefficients  are non-zero, then the  convergence rate  of the smoothed expectile MEL estimations  is slower than that of adaptive LASSO smoothed expectile MEL estimations (see Table \ref{Tab2}). The sparsity obtained by simulations confirms the theoretical result of Theorem \ref{Th3 ZW}. We proposed two algorithms for each of the estimators which turned out to be similar. As specified above, the two kept algorithms confirm the theoretical results and moreover they are simple to implement.
   \section{Application on real data}
   \label{sect_appli}
   \begin{figure*} 
   	\begin{minipage}[b]{0.45\linewidth}
   		\centering \includegraphics[scale=0.35]{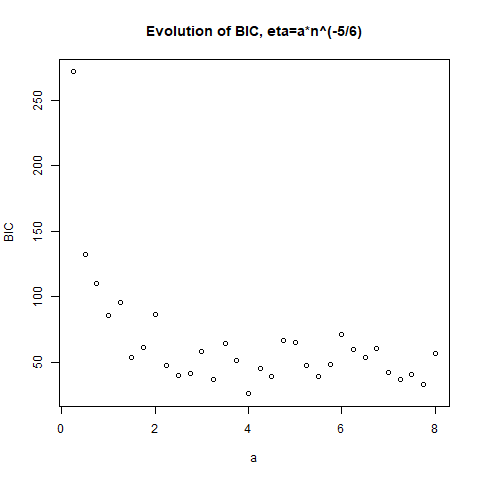}
   		\caption{\it    Evolution of BIC criterion for eyedata set  with respect to "a" for tuning sequence $\eta_n=an^{-5/6}$.}
   		\label{fig_BIC_n56}
   	\end{minipage} \hfill
   	\begin{minipage}[b]{0.45\linewidth}
   		\centering \includegraphics[scale=0.35]{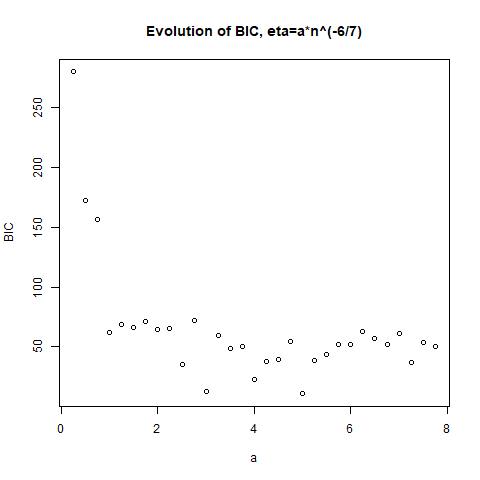}
   		\caption{ \it  Evolution of BIC criterion for eyedata set  with respect to "a" for tuning sequence $\eta_n=an^{-6/7}$.}
   		\label{fig_BIC_n67}  
   	\end{minipage} \hfill
   	   \end{figure*}
      \begin{table}
      	\caption{\footnotesize Results for \textit{eyedata} set by adaptive LASSO smoothed expectile MEL method for two tunning sequence $\eta_n$: $\eta=a n^{-5/6}$ and $\eta=a n^{-6/7}$, with $a$ found by minimizing the BIC criterion.}
      	\begin{center}
      		{\scriptsize
      			\begin{tabular}{|c|c|c|c|c|c|c|c|}\hline  
      				$\eta_n$ & $\argmin BIC(\eta_n)$  &	$\min(BIC(\eta_n))$	&  $\widehat{\cal A}^*_n$  &  $\widehat \beta^*_{n,j}$, $j \in \widehat{\cal A}^*_n$  & $\widehat{\cal R}_n^*(\widehat\eb_n^*, \el(\widehat\eb_n^*))$ & $c_{1-\alpha;dl}$ & pvalue  \\ \hline
      			$a n^{-5/6}$	 & a=4 & 25.8 & $\{8, 19,153\}$ & $\widehat \beta^*_{n,8}=0.02$ & 11.5& 7.8 & 0.009 \\
      		 	 &  &   &   &  $\widehat \beta^*_{n,19}=-0.56$, $\widehat \beta^*_{n,153}=1.01$ & &  &  \\ \hline
      		 	 $a n^{-6/7}$	 & a=5 & 11.02 & $\{19,155\}$ & $\widehat \beta^*_{n,19}=-0.54$, $\widehat \beta^*_{n,155}=0.95$ & 1.45& 5.99 & 0.48 \\ \hline
      			\end{tabular} 
      		}
      	\end{center}
      	\label{Tab3} 
      \end{table}
 In this section we will apply  the adaptive LASSO smoothed expectile MEL estimation method  to dataset  \textit{eyedata}  in the R package \textit{flare} where the response variable of gene TRIM32 is explained by 200 genes. Because the number $n$ of observations is 120, in order to have an identifiable regression model with fewer explanatory variables than observations, we consider two models, each with $p=100$ regressors. For each of the two models we select the relevant explanatory variables, after which we select the variables again for the model with these variables combined.
 In general in applications the value of $\tau$ which satisfies assumption $\eE[\rho'_\tau(\varepsilon)]=0$ is unknown because the law of the model errors  is unknown.  We then propose an empirical  method to calculate an estimator based on the explained variable values, more precisely  $\tau=\frac{n^{-1}\sum^n_{i=1} \widetilde{y}_i \e1_{\widetilde{y}_i <0}}{n^{-1} \big( \sum^n_{i=1} \widetilde{y}_i \e1_{\widetilde{y}_i <0}-\sum^n_{i=1} \widetilde{y}_i \e1_{\widetilde{y}_i >0}\big)}$, where   $\widetilde{y}_i=(y_i-median (y_1, \cdots , y_n))/\big(n^{-1} \sum^n_{i=1}|y_i-median (y_1, \cdots , y_n)|\big)$. The value of $\tau$ obtained on our data is $\tau=0.57$. The other considered parameters are $\gamma=2.5$, $h=n^{-1/4}$, $\epsilon=10^{-5}$, $\nu=0.1$, while the explanatory variables are standardized. In order to choose the optimal value of the tuning parameter $\eta_n$ we minimize the BIC criterion given by (\ref{crit_BIC}). From Figure \ref{fig_BIC_n56} we deduce that if $\eta_n$ is of the form $\eta_n=a n^{-5/6}$, then  the minimum of the BIC criterion is achieved for $a=4$.  Thus, for $\widehat \eta^*_n=4n^{-5/6}$, performing a least squares regression with the selected explanatory variables $X_8,X_{19},X_{153}$  we obtain that the coefficient of $X_{19}$ is not significant, and the residuals are gaussian. If $\eta_n$ is of the form $\eta_n=a n^{-6/7}$, then the minimum of the BIC criterion is achieved for $a=5$ (see Figure \ref{fig_BIC_n67}).  Thus, for $\widehat \eta^*_n=5n^{-6/7}$, using the   \textit{L2} estimation algorithm, only two explanatory variables have non-zero coefficients: $X_{19}, X_{155}$. Hence, for this value of $\eta_n$  the model estimation results are better, something also confirmed by the fact that the non-zero coefficients estimations   belong to the confidence region(CR) with a p-value equal to 0.48.  See all obtained results for these two $\widehat \eta^*_n$  in Table \ref{Tab3}.
\section{Proofs}
\label{sect_proof}
This section is divided in two subsections.  In the first subsection we give the proofs of the results stated in Section  \ref{section_EMVexpectile} on the smoothed expectile MEL, while in the second subsection we give the proofs of Section \ref{section_EMVaLASSO} on the adaptive estimator.
\subsection{Proofs of results in Section \ref{section_EMVexpectile}}
\label{proofs_EMVexpectile}
 {\bf Proof of   Lemma  \ref{Lemma_tl}} By the definition of the function  $G_h$ we have $G \big( \frac{\eX^\top (\eb-\ebo) -\varepsilon}{h}\big) =$ \\
 $\int^1_{-1} \e1_{x < \eX^\top (\eb-\ebo) - hv } K(v) dv$. 
In order to facilitate the   reading and understanding of the proof, let us consider the following random  variable  $V_{\eb} \equiv G \big( \frac{\eX^\top (\eb-\ebo) -\varepsilon}{h}\big) -  \e1_{\varepsilon < \eX^\top (\eb-\ebo)  }$.\\
(i)  Claim (i) is proved if we show that $V_{\ebo}  \overset{\PP_\varepsilon} {\underset{h \rightarrow 0}{\longrightarrow}}  0$. \\
For this purpose, we calculate   $\eE_\varepsilon[|V_{\ebo}|]$ which, because that the  support of kernel $K$ is $[-1,1]$ and since  $K$ is a density, is equal to: 
$$
\eE_\varepsilon[|V_{\ebo}|] = \int_{\R} \bigg| \int^1_{-1} \big( \e1_{x < - hv} - \e1_{x< 0}\big) K(v) dv\bigg| f_\varepsilon(x) dx.$$
 Using the identity
\begin{equation}
	\label{uxa}
	\e1_{x<a} -\e1_{x<0}= \textrm{sgn}(a)\e1_{\min(a,0) \leq x \leq \max (a,0)},
\end{equation} 
we get:
\begin{align*}
	\eE_\varepsilon[|V_{\ebo}|] & =\int_{\R} \bigg| \int^1_{-1} \textrm{sgn}(-v)  \e1_{h \min (0, -v) \leq  x \leq h \max (0,-v)}   K(v) dv\bigg| f_\varepsilon(x) dx \\
	& =\int_{\R} \bigg|  \int^0_{-1}  \e1_{0 \leq  x \leq - h  v} K(v) dv -  \int^1_{0}    \e1_{- h  v \leq  x \leq 0}    K(v) dv\bigg| f_\varepsilon(x) dx .
\end{align*}
Since $K$ is a kernel, we have $K(u)=K(-u)$, from where we obtain  $ \int^0_{-1}   \e1_{0 \leq  x \leq - h  v} K(v)dv = \int_0^{1}   \e1_{0 \leq  x \leq  h  v} K(v)dv $, which implies:
\begin{align}
	\eE_\varepsilon[|V_{\ebo}|] 
	& \leq  \eE_\varepsilon \bigg[     \int_0^{1}   \e1_{0 \leq  \varepsilon \leq  h  v} K(v)dv +    \int^1_{0}    \e1_{- h  v \leq  \varepsilon \leq 0}    K(v) dv \bigg] .
	\label{EV2b}
\end{align}
For the first term of the right-hand side of inequality  (\ref{EV2b}), we have: $
	\eE_\varepsilon \big[  \int_0^{1}   \e1_{0 \leq  \varepsilon \leq  h  v} K(v)dv   \big]   = \int^1_{0} \big( F_\varepsilon(hv) - F_\varepsilon(0)\big) K(v) dv  =\int^1_{0}  hv f_\varepsilon (\widetilde v) K(v) dv$,
with $\widetilde v$ between 0 and $ hv$. Because in addition the density $f_\varepsilon$ is bounded  in a neighborhood of 0 by assumption(A1)(b) and moreover $K$ is a density, we have: $ \eE_\varepsilon \big[  \int_0^{1}   \e1_{0 \leq  \varepsilon \leq  h  v} K(v)dv   \big]=O(h)$. 
We show similarly that the second term of the right-hand side of inequality  (\ref{EV2b}) is also $O(h)$, which implies that $\eE_\varepsilon[|V_{\ebo}|]=O(h)$. Claim (i) follows by the Markov inequality for $|V|$:
\begin{equation*}
	\label{ev1b}
	\PP_\varepsilon\big[ |G_h(-\varepsilon) -  \e1_{\varepsilon < 0}| \geq \epsilon\big] \leq \frac{\eE_\varepsilon[|G_h(-\varepsilon) -  \e1_{\varepsilon < 0}|]}{\epsilon}=\frac{O(h)}{\epsilon}, \qquad \forall \epsilon >0.
\end{equation*}
(ii)   Using the same arguments as in the proof of  claim (i) we have:
\begin{align}
	\eE_\varepsilon[|V_\eb|] & = \eE_\varepsilon \big[\big|  \int^1_{-1} \big( \e1_{\varepsilon - \eX^\top (\eb-\ebo) < -hv} - \e1_{\varepsilon - \eX^\top (\eb-\ebo) < 0} \big) K(v) dv \big|\big] \nonumber \\
	& = \eE_\varepsilon \big[   \big| \int^0_{-1} \e1_{0< \varepsilon - \eX^\top (\eb-\ebo)  < -hv } K(v) dv - \int^1_0 \e1_{-hv < \varepsilon - \eX^\top (\eb-\ebo)  < 0} K(v) dv\big|\big] \nonumber  \\
	& = \eE_\varepsilon \big[  \int^0_{-1} \e1_{0< \varepsilon - \eX^\top (\eb-\ebo)  < -hv } K(v) dv + \int^1_0 \e1_{-hv < \varepsilon - \eX^\top (\eb-\ebo)  < 0} K(v) dv  \big] .
	\label{ert}
\end{align}
For the first term of (\ref{ert}) we have $\eE_\varepsilon \big[  \int^0_{-1} \e1_{0< \varepsilon - \eX^\top (\eb-\ebo)  < -hv } K(v) dv  \big] = \int^0_{-1} \big(   F_\varepsilon \big( \eX^\top (\eb-\ebo)   -hv  \big) - F_\varepsilon \big(   \eX^\top (\eb-\ebo)  \big)\big) K(v) dv$.
Since $\| \eb -\ebo \| \leq C n^{-1/3}$ and the random vector $\eX$ is bounded by assumption   (A3)(b), taking also into account assumption (A1)(b), then  we have: $
\eE_\varepsilon \big[  \int^0_{-1} \e1_{0< \varepsilon - \eX^\top (\eb-\ebo)  < -hv } K(v) dv  \big] =  \int^1_0 h v f_\varepsilon \big( \eX^\top (\eb-\ebo) +\widetilde v \big) K(v) dv =O_{\PP_\varepsilon}(h|\eX)$,
with $\widetilde{v}$ between 0 and $hv$. 
We show similarly that the second term of the right-hand side of  (\ref{ert}) is   $O_{\PP_\varepsilon}(h|\eX)$, wich implies  $\eE_\varepsilon [|W| | \eX]= O_{\PP_\varepsilon}(h|\eX)$. Finally, we have for all  $\epsilon >0$, by the Markov inequality: 
\begin{align*}
	\PP_\varepsilon\big[ \big|G \bigg( \frac{\eX^\top (\eb-\ebo) -\varepsilon}{h}\bigg) -  \e1_{\varepsilon < \eX^\top (\eb-\ebo)  } \big| \geq \epsilon | \eX\big]& \leq \frac{\eE_\varepsilon\big[ \big|G \big( \frac{\eX^\top (\eb-\ebo) -\varepsilon}{h}\big) -  \e1_{\varepsilon < \eX^\top (\eb-\ebo)  } \big| | \eX\big]}{\epsilon}\\
	& = \frac{O_{\PP_\varepsilon}(h|\eX)}{\epsilon}.
\end{align*}
Claim (ii) follows taking into account assumption  (A4). 
\hspace*{\fill}$\blacksquare$  \\

\noindent {\bf Proof of   Lemma  \ref{Lemma 1}}
We first prove  relation (\ref{eq1}). Using relation (\ref{eq_tl}), assumptions  (A1)(a), (A2), using Slutsky's theorem, we have for any  $i=1, \cdots, n$:
\begin{equation*}
	\begin{split}
		\eE_\varepsilon\big[\psi_h(\eX_i,Y_i;\ebo) \varepsilon_i\big] &= \eE_\varepsilon\big[ \varepsilon_i \big( \tau + (1-2 \tau) G_h(-\varepsilon_i)\big)\big] \\
		& {\underset{h \rightarrow 0}{\longrightarrow}}  \eE_\varepsilon \big[ (\tau + (1-2 \tau )\e1_{\varepsilon < 0} ) \varepsilon\big] =  \eE_\varepsilon\big[  \rho'_\tau(\varepsilon)\big] =0.
	\end{split}
\end{equation*}
Since $\varepsilon_i$ and $\eX_i$ are independent by  assumption (A4), thus 
\begin{equation}
	\label{rep_0}
	\eE_\eX\big[ \eE_\varepsilon[\widehat{\eg}_i(\ebo)] \big] {\underset{h \rightarrow 0}{\longrightarrow}} \eE[\eg_i(\ebo)]=\oo_p.
\end{equation}
Using assumptions (A1)(a), (A2), (A3)(a), (A4) and relation (\ref{eq_tl}), we now study  the variance of $\widehat{\eg}_i(\ebo)$ with respect to  $\varepsilon$ and $\eX$:
	\begin{align}
		\Var_{\varepsilon, \eX}[\widehat{\eg}_i(\ebo)] & = \Var_\eX[\delta_i \eX_i] \Var_\varepsilon \big[\psi_h(\eX_i,Y_i; \ebo) \varepsilon_i \big]  \nonumber \\
		& {\underset{h \rightarrow 0}{\longrightarrow}}   \Var_\eX[\delta \eX] \Var_\varepsilon \big[ 2^{-1}\rho'_\tau (\varepsilon)\big] = \textbf{B}.
		\label{L5}
	\end{align}
Hence, relation (\ref{eq1}) follows by the Central Limit Theorem (CLT). We now show  relation (\ref{eq2}). Using the Law of Large Numbers (LLN), under  assumptions (A1)(a), (A2), (A3)(a), (A3)(c),  (A5), we have: 
	\[
	\frac{1}{n} \sum^{n}_{i=1} \widehat{\eg}_i(\ebo)  \widehat{\eg}_i(\ebo)^\top = 	\frac{1}{n} \sum^{n}_{i=1} \delta_i \psi^2_h(\eX_i, Y_i;\ebo) \varepsilon_i^2 \eX_i \eX_i^\top \overset{\eP} {\underset{n \rightarrow \infty}{\longrightarrow}} \textbf{B},
	\]
that is relation (\ref{eq2}). We now show relation (\ref{eq4}).
Since $K$ has a compact support on $[-1,1]$, we deduce that $\psi_h\eX_i,Y_i;\ebo)$	 takes values between 0 and 1 with a probability equal to 1. Hence, the following relations occur for any $i=1, \cdots ,n$ with probability 1: $\|\widehat{\eg}_i(\ebo) \| \leq \| \varepsilon_i \eX_i \| =|\varepsilon_i| \cdot \| \eX_i\| \leq |\varepsilon_i|\underset{1 \leqslant j \leqslant n}{\text{max}}  \| \eX_j\|  $. Moreover, by assumptions (A2), (A3)(b), (A4) we have $\eE \big[|\varepsilon_i|^2\underset{1 \leqslant j \leqslant n}{\text{max}} \| \eX_j\|^2   \big]< \infty$  and by a proof analogous to that of Lemma 3 in \cite{Owen.90} we obtain 
\begin{equation}
\label{opn}
|\varepsilon_i| \max_{1 \leqslant j \leqslant n} \| \eX_j\|=o_\PP(n^{1/2}).
\end{equation}
Thus, $  \underset{1 \leqslant i \leqslant n}{\text{max}}  \| \widehat{\eg}_i(\ebo) \|=o_\PP(n^{1/2})$ and relation (\ref{eq4}) is shown.
	\hspace*{\fill}$\blacksquare$  \\
	
\noindent {\bf Proof of  Theorem \ref{Lemma 2.1 OA}}
We recall that $\widehat{\eg}_i(\eb) = \delta_i\big( \tau - (1-2 \tau) G_h(\eX_i^\top \eb - Y_i)\big) \big(Y_i -\eX_i^\top \eb\big) \eX_i$.  
The proof will be done in two steps, the firs being necessary to the second.\\
\underline{Step I.} Before proving Theorem \ref{Lemma 2.1 OA}, we will first show  the following  three relations for all $\eb$ such that $\| \eb- \ebo \| \leq C n^{-1/3}  $, under assumptions  (A3)(b), (A3)(c):
\begin{equation}
	\label{eq1a}
	\qquad n^{-1/2} \sum^n_{i=1} \big(\widehat{\eg}_i(\eb) - \eE_{\varepsilon,\eX}[\widehat{\eg}_i(\eb)] \big) \overset{\cal L} {\underset{n \rightarrow \infty}{\longrightarrow}} {\cal N}(\oo_p,\textbf{B}),
\end{equation}
\begin{equation}
	\label{eq2a} \qquad n^{-1} \sum^n_{i=1} \widehat{\eg}_i(\eb) \widehat{\eg}_i(\eb)^\top \overset{\eP} {\underset{n \rightarrow \infty}{\longrightarrow}} \textbf{B},
\end{equation}
\begin{equation}
	\label{eq4a} \qquad \max_{1 \leqslant i \leqslant n} \| \widehat{\eg}_i(\eb)  \| =o_{\PP}(n^{1/2}).
\end{equation}
We  start with the proof of relation (\ref{eq2a}). \\
For a parameter $\eb$ such that $\| \eb - \ebo \| \leq C n^{-1/3}$ we study:
$
	\widehat{\eg}_i(\eb)  = \delta_i \big(\tau + (1- 2 \tau )  G_h(\eX_i^\top (\eb - \ebo) - \varepsilon_i) \big) \big( \varepsilon_i - \eX_i^\top (\eb - \ebo)\big) \eX_i $. 
Then, taking into account the independence between $\eX_i$ and $\varepsilon_i$ by assumption  (A4), together with the independence between   $\delta_i$  and $\varepsilon_i$ consequence of the fact that $Y_i$ is MAR, we have:
\begin{align}
\label{Eegib}
	\eE_\varepsilon \big[\widehat{\eg}_i(\eb)    \big] 
	& = \delta_i \eE_\varepsilon \big[ \big( \tau + (1-2 \tau) G \big(\frac{\eX_i^\top (\eb -\ebo) - \varepsilon_i}{h}\big)  \big) \big(\varepsilon_i  - \eX_i^\top (\eb -\ebo)  \big) \big] \eX_i.
\end{align}
On the other hand,  relation (\ref{eq_tl2}) of Lemma \ref{Lemma_tl}(ii) holds for any  $i=1, \cdots , n$ and for all $\eb$ such that $\| \eb - \ebo \| \leq C n^{-1/3}$.
Then, by relation  (\ref{eq_tl2}), we have that relation (\ref{Eegib}) converges for $h \rightarrow 0$ in respect to probability $\eP_\eX$ towards: 
\begin{align}
	&	\overset{\eP_\eX}   {\underset{h \rightarrow 0}{\longrightarrow}} \delta_i \eE_\varepsilon \big[ \big( \tau + (1-2 \tau)\e1_{\eX_i^\top (\eb - \ebo) -\varepsilon_i >0}\big) \big( \varepsilon_i - \eX_i^\top(\eb - \ebo)\big)\big] \eX_i \nonumber \\
	&  = \delta_i \left\{  \eE_\varepsilon \big[ \big( \tau + (1-2 \tau)\e1_{\eX_i^\top (\eb - \ebo) -\varepsilon_i >0}\big) \varepsilon_i \eX_i  \big]  - \eE_\varepsilon  \big[\big( \tau + (1-2 \tau)\e1_{\eX_i^\top (\eb - \ebo) -\varepsilon_i >0}\big) \eX_i^\top (\eb -\ebo) \eX_i \big]\right\} \nonumber \\
	& \equiv I_1 -I_2 .
	\label{I1I2}
\end{align}
We first study term $I_2$ of relation (\ref{I1I2}) which can be written, using assumption (A3)(b) together with $  \eE_\varepsilon \big[ \e1_{\varepsilon_i < \eX_i^\top (\eb -\ebo)} \big]=\PP_\varepsilon \big[ \eX_i^\top (\eb -\ebo) < \varepsilon_i\big]$:
\begin{equation}
	\label{I2}
I_2 =\delta_i \big( \tau + (1- 2 \tau ) (1- \eE_\varepsilon \big[ \e1_{\varepsilon_i < \eX_i^\top (\eb -\ebo)} \big]) \big) \eX_i ^\top (\eb - \ebo) \eX_i =O_{\PP_\eX}(\eb - \ebo).
\end{equation}
Now, we  study  term $I_1$ of relation (\ref{I1I2}) for which we have, using relation (\ref{uxa}):
$\eE_\varepsilon \big[ \big( \tau + (1-2 \tau)\e1_{\eX_i^\top (\eb - \ebo) -\varepsilon_i >0}\big) \varepsilon_i  \big] 	 =\int_{\R} \big( \tau +(1-2 \tau) \e1_{x < \eX_i^\top(\eb - \ebo)}\big) x f_\varepsilon (x)dx = \int_{\R} \big( \tau +(1-2 \tau)  \e1_{x < 0} \big)x f_\varepsilon (x)dx   +(1-2 \tau)  \int_{\R}  \big(\e1_{x < \eX_i^\top(\eb - \ebo)} -\e1_{x < 0}\big)x f_\varepsilon (x)dx= \eE_\varepsilon [\rho'(\varepsilon)]+(1-2 \tau) \textrm{sgn}(\eX_i^\top $ $\cdot (\eb-\ebo)) \int_{\R}   \e1_{\min \big(\eX_i^\top (\eb-\ebo), 0\big) \leq x \leq \max \big(\eX_i^\top (\eb-\ebo), 0\big)}x f_\varepsilon (x)dx$.
On the other hand, we have with probability 1: 
$$\bigg| \int_{\R}   \e1_{\min \big(\eX_i^\top (\eb-\ebo), 0\big) \leq x \leq \max \big(\eX_i^\top (\eb-\ebo), 0\big)}x f_\varepsilon (x)dx \bigg| \leq |\eX_i^\top \eb |.$$ 
Then, by assumptions  (A1)(a), (A1)(b), (A3)(c), (A4) we obtain:
\begin{equation}
\label{I1}
I_1= O_{\PP_\eX}(\eb -\ebo).
\end{equation}
Now, combining (\ref{I1I2}), (\ref{I2}), (\ref{I1}), we get
\begin{equation}
	\label{ti}
	\eE_\varepsilon[\widehat{\eg}_i(\eb)]=O_{\PP_\eX}(\eb-\ebo).
\end{equation}
Since $\eX_i$ has bounded support by assumption (A3)(a), combined with assumption  (A3)(b), the we have for relation (\ref{ti}):
\begin{equation}
	\label{tib}
	\eE_{\varepsilon, \eX} [\widehat \eg_i(\eb)]=O_\PP(\eb -\ebo).
\end{equation}
Moreover, similarly as for (\ref{L5}) we have that $\Var_{\varepsilon,\eX}[\widehat{\eg}_i(\eb)] {\underset{h \rightarrow 0}{\longrightarrow}} \textbf{B }$ and then relation (\ref{eq2a}) follows from the LLN. This also implies that (by CLT):
\[
n^{-1/2} \sum^n_{i=1} \big(\widehat{\eg}_i(\eb) -\eE_{\varepsilon,\eX}[\widehat{\eg}_i(\eb)]\big) \overset{\cal L} {\underset{n \rightarrow \infty}{\longrightarrow}} {\cal N}(\oo_p,\textbf{B}),
\]
that is relation (\ref{eq1a}).\\
The proof of relation (\ref{eq4a}) is similarly to that of  relation (\ref{eq4}) of Lemma \ref{Lemma 1}. By relation (\ref{tib}) we have: $\eE [\widehat \eg_i(\eb)]=O_\PP(n^{-1/3})$. Thus, analogously to the proof of relation (\ref{eq4}) we obtain for any $i=1, \cdots , n$ that: $\|\widehat{\eg}_i(\eb) \| \leq | \varepsilon_i - (\eb -\ebo)^\top \eX_i |   \underset{1 \leqslant j \leqslant n}{\text{max}} \| \eX_j\|$, with probability 1. By assumptions (A2), (A3)(b), (A4) we have $\eE \big[ \|\widehat{\eg}_i(\eb) \|^2 \big] < \infty$. By a proof analogous to that of Lemma 3 in \cite{Owen.90} we obtain  $  \underset{1 \leqslant j \leqslant n}{\text{max}}  \| \widehat{\eg}_j(\eb) \|=o_\PP(n^{1/2})$, that is relation (\ref{eq4a}).  \\
\underline{Step II.}  We now continue with the proof of  Theorem \ref{Lemma 2.1 OA}. In order to prove the theorem we will compare $	\widehat{\cal R}_n\big( \eb, \el(\eb)\big)$ for $\eb$ such that $\| \eb -\ebo \| \leq C n^{-1/3}$ with $	\widehat{\cal R}_n\big( \ebo, \el(\ebo)\big)$.  Firstly, we show that  $\|\el(\eb)\| =O(n^{-1/3})$ when $\| \eb -\ebo \| \leq C n^{-1/3}$. \\
For this we write $\el(\eb)$ under the form  $\el(\eb)=r \eth$, with $r \geq 0$ a scalar and  $\eth \in \R^p$ a $p$-vector   such that $\| \eth \| =1$. We multiply  with $\eth^\top $ on the left the relation in (\ref{LL}) and we get, with probability 1:
\begin{align}
	0_1 &  = \frac{1}{n} \bigg|\eth^\top  \sum^n_{i=1}\frac{\widehat{\eg}_i(\eb)}{1+\el(\eb)^\top \widehat{\eg}_i(\eb)}  \bigg|  =\frac{1}{n}  \bigg| \eth^\top \bigg(\sum^n_{i=1} \widehat{\eg}_i(\eb) - r  \sum^n_{i=1} \frac{\widehat{\eg}_i(\eb) \eth ^\top \widehat{\eg}_i(\eb)}{1+r \eth^\top \widehat{\eg}_i(\eb)} \bigg)   \bigg| \nonumber \\
	& \geq \frac{1}{n} r \eth^\top \sum^n_{i=1} \frac{\widehat{\eg}_i(\eb) \widehat{\eg}_i(\eb)^\top}{1+r \eth^\top \widehat{\eg}_i(\eb)} \eth - \frac{1}{n} \bigg| \eth^\top \sum^n_{i=1} \widehat{\eg}_i(\eb) \bigg|.
	\label{L1}
\end{align}
Since $p_i=n^{-1} (1+r \eth^\top \widehat{\eg}_i(\eb) )^{-1} \geq 0$,   we obtain $1+r \eth^\top \widehat{\eg}_i(\eb) \geq 0$ with probability 1. Moreover, by the CLT, we have: $n^{-1}\sum^n_{i=1} \big(\widehat{\eg}_i(\eb) -\eE_{\varepsilon,\eX}[\widehat{\eg}_i(\eb)]\big)  =O_\PP(n^{-1/2})$, 
from where, using relation (\ref{tib}),  we get:
\begin{equation}
	\label{L6} \frac{1}{n}\sum^n_{i=1}\widehat{\eg}_i(\eb)  -O_\PP(\eb -\ebo)=O_\PP(n^{-1/2}).
\end{equation}
On the other hand, combining the  Cauchy-Schwarz inequality together with $\| \eth \|=1$ and since each component of $n^{-1} \sum^n_{i=1}  \widehat{\eg}_i(\eb) $ is $O_\PP(n^{-1/3})$ by (\ref{L6}), we obtain for the second term on the right-hand side of (\ref{L1}):
\begin{equation}
	\label{L2}
	\frac{1}{n} \bigg| \eth^\top \sum^n_{i=1} \widehat{\eg}_i(\eb) \bigg| \leq \| \eth \|  \bigg\| \frac{1}{n} \sum^n_{i=1}  \widehat{\eg}_i(\eb)   \bigg\| =O_\PP(n^{-1/3}).
\end{equation} 
For the first term on the right-hand side of (\ref{L1}), by the Cauchy-Schwarz inequality together with  $\| \eth \|=1$, we have with probability 1:
\begin{align*}
	\frac{1}{n} r \eth^\top \sum^n_{i=1} \frac{\widehat{\eg}_i(\eb) \widehat{\eg}_i(\eb)^\top}{1+r \eth^\top \widehat{\eg}_i(\eb)} \eth & \geq \frac{1}{n} r \eth^\top \sum^n_{i=1} \frac{\widehat{\eg}_i(\eb) \widehat{\eg}_i(\eb)^\top}{1+|r \eth^\top \widehat{\eg}_i(\eb)|} \eth  \geq \frac{1}{n} r \eth^\top \sum^n_{i=1} \frac{\widehat{\eg}_i(\eb) \widehat{\eg}_i(\eb)^\top}{1+r \|\eth \| \cdot \|  \widehat{\eg}_i(\eb)\|} \eth    \\
	&  \geq \frac{1}{n} r \eth^\top \sum^n_{i=1} \frac{\widehat{\eg}_i(\eb) \widehat{\eg}_i(\eb)^\top}{1+r  \max_{1\leqslant j \leqslant n}\|  \widehat{\eg}_j(\eb)\|} \eth .
\end{align*}
On the other hand, when $\| \eb -\ebo \| \leq C n^{-1/3}$, we have  by relation (\ref{eq4a}) that  $ \underset{1 \leqslant i \leqslant n}{\text{max}} \|  \widehat{\eg}_i(\eb)\|=o_\PP(n^{1/2})$, which implies:
\begin{equation}
	\label{L3}
	\frac{1}{n} r \eth^\top \sum^n_{i=1} \frac{\widehat{\eg}_i(\eb) \widehat{\eg}_i(\eb)^\top}{1+r \eth^\top \widehat{\eg}_i(\eb)} \eth	\geq \frac{r \| \eth \|^2}{1+r  \underset{1 \leqslant j \leqslant n}{\text{max}}  \| \widehat{\eg}_j(\eb)\|} \mu_{\min} \big( \frac{1}{n}  \sum^n_{i=1} \widehat{\eg}_i(\eb) \widehat{\eg}_i(\eb)^\top \big).
\end{equation}
In relation (\ref{L3}),  $ \mu_{\min} $ is the smallest eigenvalue of the matrix, eigenvalue that, using assumption (A3)(c) and relation (\ref{eq2a}), is strictly greater than the constant  $4^{-1} C_1 \eE_\varepsilon [\rho_\rho '(\varepsilon)]^2$, with probability converging to 1. 
From relations (\ref{L1}), (\ref{L2}) and (\ref{L3}) we obtain, for $n$ large enough, with probability converging to 1, that:
\begin{equation}
	\label{rl}
	0 \geq \frac{Cr -O_\PP(n^{-1/3})}{1+r o_\PP(n^{1/2})},  \quad \textrm{from where} \quad r=\| \el(\eb) \| =O_\PP(n^{-1/3}).
\end{equation}
With these results we now show that the minimum of $\widehat {\cal R}_n (\eb, \el(\eb)) $ can only be attained in the interior of the ball $\| \eb -\ebo \|\leq C n^{-1/3}$. In relation (\ref{LL})  we denote by  $\gamma_i  =\el(\eb)^\top  \widehat{\eg}_i(\eb)$ and we use the following identity $1/(1+\gamma_i) = 1- \gamma_i +\gamma_i^2/(1+\gamma_i)$.
Hence, relation (\ref{LL}) becomes
\begin{align}
	\label{L7b}
	\oo_p &= \frac{1}{n} \sum^n_{i=1} \widehat{\eg}_i(\eb)  \bigg(1- \el(\eb)^\top  \widehat{\eg}_i(\eb) +\frac{ ( \el(\eb)^\top  \widehat{\eg}_i(\eb))^2}{1+\el(\eb)^\top  \widehat{\eg}_i(\eb)} \bigg). 
\end{align}
But, $1+\el(\eb)^\top  \widehat{\eg}_i(\eb)> 0$ holds with probability 1 since $p_i>0$. Applying the  Cauchy-Schwarz inequality  together with  relation of (\ref{rl}), we obtain:
\begin{equation}
	\label{ri}
\begin{split}
	\frac{1}{n}\bigg\| \sum^n_{i=1} \widehat{\eg}_i(\eb) \frac{\big(\widehat{\eg}_i(\eb)^\top \el(\eb) \big)^2}{1+\widehat{\eg}_i(\eb)^\top \el(\eb)}  \bigg\| &\leq \frac{1}{n} \sum^n_{i=1} \|  \widehat{\eg}_i(\eb)\|^3\frac{\| \el(\eb) \|^2}{ 1+\widehat{\eg}_i(\eb)^\top \el(\eb) }\\
	& =O_\PP\big(n^{-2/3}\big)\frac{1}{n} \sum^n_{i=1} \frac{ \|  \widehat{\eg}_i(\eb)\|^3}{ 1+\widehat{\eg}_i(\eb)^\top \el(\eb) }.
	\end{split}
\end{equation} 
By relation (\ref{rep_0}), we have that $\eE_{\varepsilon, \eX} \big[\widehat{\eg}_i(\ebo) \big]{\underset{h \rightarrow 0}{\longrightarrow}} \eE_{\varepsilon, \eX}\big[ {\eg}_i(\ebo)\big]=\oo_p$ and $\eE_{\varepsilon, \eX} \big[\widehat{\eg}_i(\eb) \big]=O(\eb -\ebo)$ by relation (\ref{tib}). Similar as for relations (\ref{L5}) and (\ref{tib}), we can prove  $\Var\big[\widehat{\eg}_i(\eb) \big]=O(\eb -\ebo)$. By the Bienaymé-Tchebychev inequality, for all $\epsilon_1 >0$ we have: $
\PP \big[\big|  \widehat{\eg}_i(\eb) -  \eE_{\varepsilon, \eX} \big[\widehat{\eg}_i(\eb) \big] \big| \geq \epsilon_1\big] \leq \epsilon_1^{-2}\Var\big[\widehat{\eg}_i(\eb) \big]$, 
from where $  \widehat{\eg}_i(\eb) =O_\PP(\eb -\ebo)$ for $ \eb -\ebo=O(n^{-1/3})$. Using relation (\ref{rl}), we obtain: $\big| \widehat{\eg}_i(\eb)^\top \el(\eb)\big| \leq \|\widehat{\eg}_i(\eb)\| \cdot \|\el(\eb) \| =O_\PP(n^{-2/3})$. This implies that there exist two positive constants $M_1, M_2 >0$ such that
\begin{equation}
	\label{M1M2}
	0 < M_1 < \big(1 +\widehat{\eg}_i(\eb)^\top \el(\eb)\big)^{-1} < M2 < \infty,
\end{equation}
with probability converging to 1. 
Taking into account this last relation together with relation (\ref{ri}), we can deduce:
\begin{equation}
	\label{L8}
	\frac{1}{n} \bigg\|\sum^n_{i=1} \widehat{\eg}_i(\eb) \frac{\big(\widehat{\eg}_i(\eb)^\top \el(\eb) \big)^2}{1+\widehat{\eg}_i(\eb)^\top \el(\eb)}   \bigg\| \leq O_\PP(n^{-2/3}) 	\frac{1}{n} \sum^n_{i=1}  \|  \widehat{\eg}_i(\eb)\|^3 =O_\PP(n^{-2/3}).
\end{equation}
For the last equality we used the assumption that $\eE_{\varepsilon, \eX}\big[  \|  \widehat{\eg}_i(\eb)\|^3 \big]$ is bounded for all  $\eb$ such that  $\| \eb -\ebo \| \leq C n^{-1/3}$, which results from assumptions  (A3)(a),  (A1)(a), (A4).
Then,  relations (\ref{L7b}) and (\ref{L8}) imply:
\begin{equation}
	\label{L9}
	\oo_p = \frac{1}{n} \sum^n_{i=1} \widehat{\eg}_i(\eb) - \frac{\el(\eb)^\top}{n}  \sum^n_{i=1} \widehat{\eg}_i(\eb) \widehat{\eg}_i(\eb)^\top +O_\PP(n^{-2/3}),
\end{equation}
relation for which we have $n^{-1} \sum^n_{i=1} \widehat{\eg}_i(\eb) =O_\PP(\eb - \ebo)=O_\PP(n^{-1/3})$ by relation (\ref{L6}) and  $ n^{-1}\el(\eb)^\top   \sum^n_{i=1} \widehat{\eg}_i(\eb) \widehat{\eg}_i(\eb)^\top=O_\PP(\el(\eb))=O_\PP(n^{-1/3})$ by relations (\ref{eq2a}) and (\ref{rl}). Relation (\ref{L9})  implies, for $\| \eb - \ebo \| \leq C n^{-1/3}$, that
\begin{equation}
	\label{L10}
	\el(\eb)=\bigg(\frac{1}{n}  \sum^n_{i=1} \widehat{\eg}_i(\eb) \widehat{\eg}_i(\eb)^\top   \bigg)^{-1}   \bigg( \frac{1}{n} \sum^n_{i=1} \widehat{\eg}_i(\eb)  \bigg) +O_\PP(n^{-2/3})=O_\PP(n^{-1/3}).
\end{equation}
For $	\widehat{\cal R}_n\big( \eb, \el(\eb)\big)$ given by relation (\ref{Rnbl}), using a Taylor expansion around $\oo_p$ with respect to $\el(\eb)$ we can write:
\begin{equation}
	\label{LU}
	\widehat{\cal R}_n\big( \eb, \el(\eb)\big)=  2  \sum^n_{i=1} \el(\eb)^\top\widehat{\eg}_i(\eb) -  \sum^n_{i=1} \big(\el(\eb)^\top\widehat{\eg}_i(\eb)\big)^2   +o_\PP(n^{1/3}) .
\end{equation}
We replace (\ref{L10}) in (\ref{LU}) and we obtain:
\begin{equation}
	\label{LUb}
	\widehat{\cal R}_n\big( \eb, \el(\eb)\big)= n \bigg(\frac{1}{n} \sum^n_{i=1} \widehat \eg_i(\eb)\bigg)^\top \bigg(\frac{1}{n} \sum^n_{i=1} \widehat \eg_i(\eb)   \widehat \eg_i^\top(\eb)  \bigg)^{-1}  \bigg(\frac{1}{n} \sum^n_{i=1} \widehat \eg_i(\eb)\bigg) +o_\PP(n^{1/3}) .
\end{equation}
Where upon, in relation (\ref{LUb}), we write $	\widehat{\eg}_i(\eb)$ in respect to $\widehat{\eg}_i(\ebo)$ taking into account their definitions: $	\widehat{\eg}_i(\eb) -\widehat{\eg}_i(\ebo)  =\delta_i \eX_i \{\big( ( \tau + (1-2 \tau)G_h(\eX_i^\top \eb -Y_i)) 
( \varepsilon_i - \eX_i^\top (\eb - \ebo) )\big) 	- \big(  \big(\tau + (1- 2 \tau)G_h(-\varepsilon_i)\big) \varepsilon_i\big)\}$.
On the other hand, considering  $\eb=\ebo+n^{-1/3} \eu$, with $\eu \in \R^p$, $\| \eu \| \leq C$ and using the Taylor expansion up tu order 1 about $\eb=\ebo$ with the  rest ${\cal \textbf{T}}_n$ Taylor-Lagrange, similarly as in \cite{Qin-Law.94}, we get for   relation (\ref{LUb}):
\begin{equation}
	\label{RgT}
	\begin{split}
		\widehat{\cal R}_n\big( \eb, \el(\eb)\big)	& =n \bigg(\frac{1}{n} \sum^n_{i=1} \widehat{\eg}_i(\ebo) + \frac{1}{n} \sum^n_{i=1} \frac{\partial  \widehat{\eg}_i(\ebo) }{\partial \eb} \eu  n^{-1/3}  +\textbf{{\cal T}}_n \bigg)^\top \bigg( \frac{1}{n} \sum^n_{i=1} \widehat{\eg}_i(\eb) \widehat{\eg}_i^\top(\eb)\bigg)^{-1}   \\
		& \qquad  \cdot \bigg(\frac{1}{n} \sum^n_{i=1} \widehat{\eg}_i(\ebo) + \frac{1}{n} \sum^n_{i=1} \frac{\partial  \widehat{\eg}_i(\ebo) }{\partial \eb} \eu n^{-1/3}+\textbf{{\cal T}}_n   \bigg) +o_\PP(n^{1/3}),
	\end{split}
\end{equation}
with $ \textbf{{\cal T}}_n $ the $p$-vector: $2^{-1} \sum^p_{k=1} \sum^p_{l=1} \frac{\partial^2 \widehat{\eg}_i( \widetilde \eb_{i,kl})}{  \partial \beta_k \partial \beta_l }(\beta_k-\beta^0_k)(\beta_l-\beta^0_l)$, with $\frac{\partial^2 \widehat{\eg}_i( \widetilde \eb_{i,kl})}{  \partial \beta_k \partial \beta_l }$ a $p$-vector with the components: $\frac{\partial^2 \widehat{\eg}_{i,j} ( \widetilde \eb_{i,kl}) }{\partial \beta_k \partial \beta_l }$, $j=1, \cdots , p$ and the random $p$-vector  $\widetilde \eb_{i,kl}=\ebo +a_{i,kl}(\eb -\ebo)$, with $a_{i,kl} \in [0,1]$. It was denoted by $\widehat{g}_{i,j}$ the $j$-th component of the vector $\widehat{\eg}_i$. We also point out that $\frac{\partial  \widehat{\eg}_i(\ebo) }{\partial \eb}$ is the Jacobian of $p$-vector $  \widehat{\eg}_i(\eb)$ on point $\ebo$. \\
By relation (\ref{eq1}) of Lemma \ref{Lemma 1},  we have:
\begin{equation}
	\label{L22}
	\frac{1}{n} \sum^n_{i=1} \widehat{\eg}_i(\ebo)=O_\PP(n^{-1/2}).
\end{equation}
We now  study  $n^{-1} \sum^n_{i=1}\frac{\partial  \widehat{\eg}_i(\ebo)}{\partial \eb} \eu n^{-1/3}$ of relation (\ref{RgT}). 
For the square matrix  $\frac{\partial  \widehat{\eg}_i(\ebo) }{\partial \eb}$ of dimension $p \times p$ we have: $	\frac{\partial  \widehat{\eg}_i(\ebo) }{\partial \eb}  =\delta_i\eX_i  \big( (1-2 \tau) \frac{\partial G_h(\eX_i^\top \eb -Y_i)}{\partial \eb} \eX_i^\top (Y_i - \eX_i^\top \eb) -\eX_i^\top \big( \tau + (1-2 \tau) G_h(\eX_i^\top \eb- Y_i) \big)  \big)_{|{\eb =\ebo}}$, from where
\begin{equation}
	\frac{\partial  \widehat{\eg}_i(\ebo) }{\partial \eb}  = \delta_i\eX_i \eX_i^\top \big( (1-2 \tau) \frac{\partial G_h(-\varepsilon_i)}{\partial \eb}   \varepsilon_i -  \big( \tau + (1-2 \tau) G_h(-\varepsilon_i) \big)  \big).
	\label{ce}
\end{equation}
On the other hand, since $G'=K$, we have $
\frac{\partial G_h (\eX_i^\top \eb - Y_i)}{\partial \eb} =\frac{\eX_i^\top}{h} K \big(\frac{\eX_i^\top (\eb - \ebo) - \varepsilon_i}{h}\big)$. Thus, relation (\ref{ce})  becomes $
\frac{\partial  \widehat{\eg}_i(\ebo) }{\partial \eb}= \delta_i\eX_i \eX_i^\top \big((1-2 \tau)  K (-  {\varepsilon_i}/{h}) \varepsilon_i/h - \big(\tau +(1-2 \tau) G(- {\varepsilon_i}/{h})\big) \big)$. 
Therefore, for the second term of (\ref{ce}) we have: 
$
n^{-1} \sum^n_{i=1}\delta_i\eX_i \eX_i^\top \big(\tau + (1-2 \tau) G_h(-\varepsilon_i) \big)=n^{-1} \sum^n_{i=1}\delta_i\eX_i \eX_i^\top \big(\tau + (1-2 \tau) \big(G_h(-\varepsilon_i) -\e1_{\varepsilon_i <0} +\e1_{\varepsilon_i <0}\big)\big)$.
By the  LLN, using assumptions (A1)(a)  and (A4), we have:
$n^{-1} \sum^n_{i=1}\delta_i\eX_i \eX_i^\top \big(\tau + (1-2 \tau)  \e1_{\varepsilon_i <0} \big) \overset{\PP} {\underset{n \rightarrow \infty}{\longrightarrow}} \eE_{\varepsilon, \eX}\big[\delta\eX \eX^\top  \big( \tau +(1 - 2 \tau) \e1_{\varepsilon <0}\big)\big] =  \eE_\eX\big[\delta\eX \eX^\top  \big] \eE_\varepsilon \big[\rho'_\tau(\varepsilon)\big]= 0$,
which implies, by CLT: 
\[
\frac{1}{n} \sum^n_{i=1}\delta_i\eX_i \eX_i^\top \big(\tau + (1-2 \tau)  \e1_{\varepsilon_i <0} \big)=O_\PP(n^{-1/2}).
\]
On the other hand, by the LLN and the proof of  Lemma \ref{Lemma_tl}(i), we have:
$n^{-1} \sum^n_{i=1}\delta_i\eX_i \eX_i^\top \big(\tau + (1-2 \tau) \big(G_h(-\varepsilon_i) -\e1_{\varepsilon_i <0}  \big)\big) \overset{\PP} {\underset{n \rightarrow \infty}{\longrightarrow}} \tau \eE_\eX\big[\delta\eX \eX^\top  \big]+(1-2 \tau) \eE_\eX\big[\delta\eX \eX^\top  \big] \eE_\varepsilon[G_h(-\varepsilon) -\e1_{\varepsilon <0}]	=\tau \eE_\eX\big[\delta\eX \eX^\top  \big]+(1-2 \tau) \eE_\eX\big[\delta\eX \eX^\top  \big] O(h)=O(1)$.  Then, we get for the second term of (\ref{ce}):
\begin{equation}
	\label{cea}
	\frac{1}{n} \sum^n_{i=1}\delta_i\eX_i \eX_i^\top \big(\tau + (1-2 \tau) G_h(-\varepsilon_i) \big)=O_{\PP}(1).
\end{equation}
Moreover, for the first  term of (\ref{ce}), by the LLN, we have:
\begin{align*}
	\frac{1}{n} \sum^n_{i=1} K \bigg(- \frac{\varepsilon_i}{h}\bigg) \frac{\varepsilon_i}{h}  \overset{\PP_\varepsilon} {\underset{n \rightarrow \infty}{\longrightarrow}} \int^h_{-h}  K\bigg( - \frac{x}{h}\bigg)\frac{ x}{h}f_\varepsilon (x)dx =h \int^1_{-1} K(-y) y f_\varepsilon (hy) dy .
\end{align*}
We use the  supposition that $K$ is bounded by assumption  (A5) together with the fact that  $f_\varepsilon$ is bounded in a neighborhood of 0 by  assumption (A1)(b) and then we obtain that  
\begin{equation}
	\label{ceb}
	h \bigg|\int^h_{-h}  K\bigg( - \frac{x}{h}\bigg)\frac{ x}{h}f_\varepsilon (x)dx\bigg| =h \bigg| \int^1_{-1} K(-y) y f_\varepsilon (hy) dy  \bigg|  \leq C h.
\end{equation}
Combining  relations (\ref{ce}), (\ref{cea}), (\ref{ceb}) we obtain $n^{-1} \| \sum^n_{i=1} \frac{\partial \widehat{\eg}_i(\ebo)}{\partial \eb} \eu n^{-1/3}\| = O_\PP(n^{-1/3})$. Then, taking into account  relation (\ref{L22}), we have:    
\begin{equation}
	\label{L33b}
	n^{-1} \| \sum^n_{i=1} \widehat{\eg}_i(\ebo) \|  =O_\PP(n^{-1/2}) \ll n^{-1} \| \sum^n_{i=1} \frac{\partial \widehat{\eg}_i(\ebo)}{\partial \eb} \eu n^{-1/3}\| = O_\PP(n^{-1/3}).
\end{equation}
For each component ${\cal T}_{n,j}$ of $\textbf{\cal T}_n$ of relation (\ref{RgT}), for $j=1, \cdots, p$, we consider the notation $\frac{\partial^2 \widehat{\eg}_{i,j}(  \eb)  }{\partial \eb^2} \equiv \bigg(  \frac{\partial^2 \widehat{\eg}_{i,j}(  \eb)}{  \partial \beta_k \partial \beta_l } \bigg)_{1 \leqslant l, k \leqslant p}$ which is a square matrix of dimension $p \times p$. By elementary calculations, we obtain
\begin{equation}
\begin{split}
	\frac{\partial^2 \widehat{\eg}_{i,j}(  \eb)  }{\partial \eb^2}
	 = \delta_i X_{i,j}  \bigg(    (1-2 \tau) \eX_i \frac{1}{h^2}K' \bigg(  \frac{\eX_i^\top (\eb - \ebo)- \varepsilon_i}{h}\bigg) \eX_i^\top (Y_i - \eX_i^\top \eb) 
	 & \\-2(1-2 \tau)\eX_i \frac{1}{h}K \bigg(  \frac{\eX_i^\top (\eb - \ebo)- \varepsilon_i}{h}\bigg) \eX_i^\top   \bigg).
	 \end{split}
	\label{TT}
\end{equation}
Since $K'$ is bounded by assumption (A5), using assumptions (A1)(a) and (A3)(b), we have, the inequalities being with probability one:
\begin{align*}
	\frac{1}{n} \bigg\|  \sum^n_{i=1} \delta_i X_{i,j} \frac{\eX_i \eX_i^\top}{h^2}K' \bigg(  \frac{\eX_i^\top (\eb - \ebo)- \varepsilon_i}{h}  \bigg) \varepsilon_i \bigg\| & \leq \frac{1}{n} \sum^n_{i=1} \delta_i  \bigg\| X_{i,j} \frac{\eX_i \eX_i^\top}{h^2}K' \bigg(  \frac{\eX_i^\top (\eb - \ebo)- \varepsilon_i}{h} \bigg) \varepsilon_i  \bigg\| \\
	& \leq C \frac{1}{n h^2} \sum^n_{i=1} | \varepsilon_i |=O_\PP (h^{-2}).
\end{align*}
Thus, using this last relation combined with assumptions (A3)(b), (A5), we have
\[
\frac{1}{n} \sum^n_{i=1} \delta_i X_{i,j} \frac{\eX_i \eX_i^\top}{h^2}K' \bigg(  \frac{\eX_i^\top (\eb - \ebo)- \varepsilon_i}{h} \bigg)   \big(\varepsilon_i -\eX_i^\top  (\eb -\ebo)\big) =O_\PP(h^{-2}+n^{-1/3} h^{-2})=O_\PP(h^{-2}).
\]
For the last term of (\ref{TT}), using also assumption (A3)(b), we have:
\[
\bigg\| \frac{1}{n} \sum^n_{i=1} \delta_i X_{i,j}  \eX_i \frac{\eX_i^\top}{h}K \bigg(  \frac{\eX_i^\top (\eb - \ebo)- \varepsilon_i}{h}\bigg)\bigg\|  \leq \frac{C}{n} \frac{1}{h}  \sum^n_{i=1} K \bigg(  \frac{\eX_i^\top (\eb - \ebo)- \varepsilon_i}{h}\bigg).
\]
Since $K$ is bounded by assumption  (A5), we deduce:
\[
\bigg\| \frac{1}{n} \sum^n_{i=1} \delta_i X_{i,j}  \eX_i \frac{\eX_i^\top}{h}K \bigg(  \frac{\eX_i^\top (\eb - \ebo)- \varepsilon_i}{h}\bigg)\bigg\|  \leq  O_\PP(h^{-1}).
\]
Then, the rest  $\textbf{{\cal T}}_n$ of  relation (\ref{RgT}) is  $O_\PP(n^{-2/3} (h^{-2}+h^{-1}))=O_\PP(n^{-2/3}h^{-1})$. Taking into account  relation (\ref{L33b}) together with  $n^{1/3}h \rightarrow\infty$, we obtain that  $\textbf{{\cal T}}_n$ is much smaller than $n^{-1} \big\|\sum^n_{i=1} \frac{\partial \widehat{\eg}_i(\ebo)}{\partial \eb} \eu n^{-1/3} \big\|$, with a probability converging to 1.\\
Hence, in $ \widehat{\cal R}_n\big( \eb, \el(\eb)\big)$ of relation (\ref{RgT}), for $\| \eb - \ebo \| \leq C n^{-1/3}$, the following term dominates:
\begin{equation}
	\label{L11}
	n \bigg(  \frac{1}{n} n^{-1/3} \sum^n_{i=1} \frac{\partial  \widehat{\eg}_i(\ebo)  }{\partial \eb} \eu\bigg)^\top \bigg( \frac{1}{n}  \sum^n_{i=1}  \widehat{\eg}_i(\ebo) \widehat{\eg}_i(\ebo)^\top  \bigg)^{-1}\bigg(  \frac{1}{n} n^{-1/3} \sum^n_{i=1} \frac{\partial  \widehat{\eg}_i(\ebo)  }{\partial \eb} \eu\bigg) \geq C n^{1/3}.
\end{equation} 
We now focus on the asymptotic study of $ \widehat{\cal R}_n\big( \ebo, \el(\ebo)\big)$. First,  similarly to the proof of Theorem 2 of \cite{Zhang-Wang.20}, we will prove that   $\| \el(\ebo) \| =O_\PP(n^{-1/2})$. For this, we use  relation (\ref{L1}) for $\eb=\ebo$ together with the  Cauchy-Schwarz inequality, the fact that  $\| \eth \|=1$ and we obtain: 
\begin{align}
	\frac{1}{n} \bigg| \eth^\top \sum^n_{i=1} \widehat{\eg}_i(\ebo) \bigg| & \geq \frac{1}{n}  r \eth^\top \sum^n_{i=1} \frac{\widehat{\eg}_i(\ebo) \widehat{\eg}_i(\ebo)^\top}{1+r \eth^\top \widehat{\eg}_i(\ebo)} \eth \nonumber \\
	& \geq \frac{r}{1+r \underset{1 \leqslant j \leqslant n}{\text{max}} \| \widehat{\eg}_j(\ebo)\| } \frac{\eth^\top}{n} \sum^n_{i=1} \widehat{\eg}_i(\ebo) \widehat{\eg}_i(\ebo)^\top \eth .
	\label{rep}
\end{align}
Combining  relation (\ref{L22}) and the  Cauchy-Schwarz inequality, we obtain: $
n^{-1} \big| \eth^\top \sum^n_{i=1} \widehat{\eg}_i(\ebo) \big| \leq \| \eth \| \big\| n^{-1} \sum^n_{i=1} \widehat{\eg}_i(\ebo)\big\| =O_\PP(n^{-1/2})$. From relation (\ref{L3}), we have using relation (\ref{eq2}) the following: $n^{-1}  r \eth^\top \sum^n_{i=1} \widehat{\eg}_i(\ebo) \widehat{\eg}_i(\ebo)^\top \eth \geq r \| \eth \|^2 \mu_{\min}( \textbf{B}) \big(1+o_\PP(1)\big)$.
Therefore, combining this last relation together with (\ref{eq4}) we obtain that  relation (\ref{rep})  becomes: $O_\PP(n^{-1/2}) \geq r \big(1+r o_\PP(n^{-1/2})\big)^{-1} $ and then  $r=O_\PP(n^{-1/2})$, that is  $\el(\ebo)=O_\PP(n^{-1/2})$.\\
Hence, applying the Cauchy-Schwarz inequality and involving  relation (\ref{eq4}), we have:
\[
\max_{1 \leqslant i \leqslant n} \big| \el^\top(\ebo) \widehat{\eg}_i(\ebo)\big| \leq \| \el(\ebo)\| \max_{1 \leqslant i \leqslant n} \| \widehat{\eg}_i(\ebo) \|=O_\PP(n^{-1/2})o_\PP(n^{1/2})=o_\PP(1).
\]
On the other hand, for $\eb=\ebo$, relation (\ref{L10}) becomes:
\begin{equation}
	\label{ZW1}
	\el(\ebo)=\bigg( \frac{1}{n} \sum^n_{i=1} \widehat{\eg}_i(\ebo)  \widehat{\eg}_i(\ebo)^\top  \bigg)^{-1} \bigg( \frac{1}{n} \sum^n_{i=1} \widehat{\eg}_i(\ebo)\bigg)+o_\PP(n^{-1/2}).
\end{equation}
Taking into account  relation (\ref{ZW1}), by similar calculations given in the proof of Theorem 2 of \cite{Zhang-Wang.20}, we obtain:
\begin{align}
	\label{ZW2}
	\widehat{\cal R}_n(\ebo, \el(\ebo))&=2 \el^\top(\ebo) \sum^n_{i=1} \widehat{\eg}_i(\ebo)- \el^\top(\ebo) \sum^n_{i=1} \widehat{\eg}_i(\ebo)  \widehat{\eg}_i(\ebo)^\top \el^\top(\ebo)  +o_\PP(1) \nonumber \\
	& = \bigg(\frac{1}{\sqrt{n}} \sum^n_{i=1} \widehat{\eg}_i(\ebo)^\top\bigg) \bigg( \frac{1}{n} \sum^n_{i=1} \widehat{\eg}_i(\ebo)  \widehat{\eg}_i(\ebo)^\top \bigg)^{-1} \bigg(\frac{1}{\sqrt{n}} \sum^n_{i=1} \widehat{\eg}_i(\ebo)\bigg) +o_\PP(1).
\end{align}
This last relation can be written  $\widehat{\cal R}_n\big( \ebo, \el(\ebo)\big)  = n \big(n^{-1} \sum^n_{i=1}  \widehat{\eg}_i(\ebo) \big)^\top \big(n^{-1} \sum^n_{i=1}  \widehat{\eg}_i(\ebo) $  $\cdot\widehat{\eg}_i(\ebo)^\top \big)^{-1} \big(n^{-1} \sum^n_{i=1}  \widehat{\eg}_i(\ebo) \big) $ $ \big(1+o_\PP(1)\big)$ which is, taking into account relations (\ref{L22}) and  (\ref{eq2}), equal to
\begin{equation}
	\label{L12}
\widehat{\cal R}_n\big( \ebo, \el(\ebo)\big)	=n O_\PP(n^{-1/2}) O_\PP(n^{-1/2})=O_\PP(1).
\end{equation}
Since $ \widehat{\cal R}_n\big( \eb, \el(\eb)\big)$ is a continuous function in $\eb$ in the ball $\| \eb - \ebo \| \leq C n^{-1/3}$,   relations (\ref{L11}) and (\ref{L12}) imply that the  minimum of $ \widehat{\cal R}_n\big( \eb, \el(\eb)\big)$ is inside this ball and the statement of the theorem follows.
\hspace*{\fill}$\blacksquare$  \\

\noindent {\bf Proof of  Theorem  \ref{Th1 ZW}}
Using Theorem \ref{Lemma 2.1 OA}, the proof of this theorem is similar to that of  Theorem 2.1 in  \cite{Ozdemir-Arslan.2021}  and of Theorem 1 in \cite{Qin-Law.94}.  The details are omitted.
\hspace*{\fill}$\blacksquare$  \\

\noindent {\bf Proof of  Theorem  \ref{Th2 ZW}}
By relation (\ref{L22}), we have that the convergence rate of  $n^{-1}\sum^n_{i=1} \widehat{\eg}_i(\ebo)$ is of order  $n^{-1/2}$. 
The Theorem results by taking into account  relations  (\ref{eq1}), (\ref{eq2}) and (\ref{ZW2}).
\hspace*{\fill}$\blacksquare$  

\subsection{Proofs of results in Section \ref{section_EMVaLASSO}}
\label{proofs_EMVaLASSO}
{\bf Proof of  Theorem  \ref{thh}}
Let us study the penalty of $\widehat {\cal R}^*_n(\eb, \el(\eb))$, first for $j \in {\cal A}$ and afterwards for $ j \not \in {\cal A}$. We decompose the penalty: $n \eta_n \sum^p_{j=1} \widehat \omega_{n,j}|\beta_j|=n \eta_n \sum^p_{j=1} \widehat \omega_{n,j} \big[ |\beta_j| - |\beta^0_j|\big] +n \eta_n \sum^q_{j=1} \widehat \omega_{n,j} |\beta^0_j|$.\\
\underline{If $j \in {\cal A}$}, since $\widehat \eb_n \overset{\eP}{\underset{n \rightarrow \infty}{\longrightarrow}} \ebo $, then $\widehat\omega_{n,j}=O_\PP(1)$. Thus, using  $n^{1/2} \eta_n{\underset{n \rightarrow \infty}{\longrightarrow}}  0$ which is a consequence of assumption (A6), we obtain:
\begin{align}
	n \eta_n  \widehat \omega_{n,j} \big( |\beta_j| - |\beta^0_j|\big)& = O_\PP\big( n \eta_n \big(|\beta_j|-|\beta^0_j|\big)\big) =n \eta_n O_\PP(n^{-1/3})  \nonumber \\
	&=O_\PP\big( (n^{1/2} \eta_n) n^{1/6}\big)=o_\PP(n^{1/6})=o_\PP(n^{1/3}).
	\label{ups1}
\end{align}
\underline{If $ j \not \in {\cal A}$}, then $
\widehat \omega_{n,j}= |\widehat \beta_{n,j}|^{-\gamma}=\ O_\PP \big(\big|\beta^0_j+n^{-1/3} u_j\big|^{-\gamma}\big)= O_\PP(n^{\gamma/3})$, which implies, using  $n^{2/3} \eta_n {\underset{n \rightarrow \infty}{\longrightarrow}} 0$ and that   $\gamma \leq 3$:
\begin{align}
	n \eta_n \widehat \omega_{n,j} \big(|\beta_j|-|\beta^0_j|\big)  =n \eta_n \widehat \omega_j  n^{-1/3} u_j
	= O_\PP \big(n \eta_n n^{(\gamma - 1)/3}\big) 
	= O_\PP \big(n \eta_n \big)=o_\PP(n^{1/3}).
	\label{ups2}
\end{align} 
Then, taking into account relations (\ref{Rob}), (\ref{ups1}) and (\ref{ups2}),  together with the supposition $n^{2/3} \eta_n {\underset{n \rightarrow \infty}{\longrightarrow}} 0$ of (A6), by  a similar approach to that made for  relation (\ref{L11}) we obtain, for $\|\eb -\ebo \| \leq C n^{-1/3}$, that:
\begin{align}
	\label{L11b}
	\widehat{\cal R}^*_n(\eb, \el(\eb)) & = n\bigg(\frac{1}{n} n^{-1/3} \sum^n_{i=1}  \frac{\partial \widehat{\eg}_i(\ebo)}{\partial \eb} \eu\bigg)^\top \bigg( \frac{1}{n} \sum^n_{i=1} \widehat{\eg}_i(\ebo)  \widehat{\eg}_i(\ebo)^\top \bigg)^{-1}\bigg(\frac{1}{n} n^{-1/3} \sum^n_{i=1}  \frac{\partial \widehat{\eg}_i(\ebo)}{\partial \eb} \eu\bigg) \nonumber\\
	& \qquad + n \eta_n \sum^q_{j=1} \widehat \omega_{n,j} |\beta^0_j| +o_\PP(n^{1/3}) \nonumber \\
	& \geq C n^{1/3} +n^{1/3}(n^{2/3} \eta_n) \sum^q_{j=1} \widehat \omega_{n,j} |\beta^0_j|+o_\PP(n^{1/3})=C n^{1/3}+o_\PP(n^{1/3}) .
\end{align}
On the other hand,  by relation (\ref{L12}) we have:
\begin{equation}
	\label{L11ba}
	\widehat{\cal R}^*_n(\ebo, \el(\ebo)) =O_\PP(1)+n \eta_n \sum^q_{j=1} \widehat \omega_{n,j} |\beta^0_j| =O_\PP(1)+o_\PP(n^{1/3}).
\end{equation}
From relations (\ref{L11b}) and (\ref{L11ba}) results that the minimum of $\widehat{\cal R}^*_n(\eb, \el(\eb))$ is realized for a parameter $\eb$ such that $\| \eb - \ebo\| \leq C n^{-1/3}$.
\hspace*{\fill}$\blacksquare$  \\
 
\noindent {\bf Proof of  Theorem  \ref{Th3 ZW}}
The proof of the oracle properties is based on the fact that the random $p$-vectors $\widehat \eb_n^*$ and $\el(\widehat \eb_n^*)$ are the solutions of the system of equations:
\begin{equation}
	\left\{  
	\begin{split}
		\textbf{S}_{1}^*	(\eb, \el(\eb)) \equiv 	\frac{\partial \widehat{\cal R}^*_n(\eb, \el(\eb)) }{\partial \el}=\oo_p, \\
		\textbf{S}_{2}^*	(\eb, \el(\eb)) \equiv	\frac{\partial \widehat{\cal R}^*_n(\eb, \el(\eb)) }{\partial \eb}=\oo_p.
	\end{split}
	\right. 
	\label{dRb}
\end{equation}
\underline{(i) The sparsity.} 
Note that $\textbf{S}_{1}^*(\eb, \el(\eb))=\textbf{S}_1(\eb, \el(\eb))=n^{-1}\frac{\partial \widehat{\cal R}_n(\eb, \el(\eb)) }{\partial \el}$, with $\textbf{S}_1(\eb, \el(\eb))$ and $\textbf{S}_2(\eb, \el(\eb))$ defined in the statement of Theorem \ref{Lemma 2.1 OA}.\\
For any $j=1, \cdots , p$, we have for the components of $\textbf{S}_{2}^*=n^{-1} \bigg(\frac{\partial \widehat{\cal R}^*_n}{\partial \beta_1}, \cdots , \frac{\partial \widehat{\cal R}^*_n}{\partial \beta_p}\bigg)$ in (\ref{dRb}):
\begin{align}
	\label{TS0}
	0&=S_{2,j}^*(\widehat \eb_n^*,\el(\widehat \eb_n^*)) = \frac{1}{n} \bigg( 2 \sum^n_{i=1}   \frac{\el(\widehat \eb_n^*)\frac{\partial \widehat \eg_i(\widehat \eb_n^*)}{\partial \beta_j} }{1+\el(\widehat \eb_n^*)^\top \cdot \widehat \eg_i(\widehat \eb_n^*)} +n \eta_n \widehat \omega_{n,j} \textrm{sgn}(\widehat \beta_{n,j}^*)\bigg) \\
	& =  {S}_{2,j}\big(\widehat \eb_n^*, \el(\widehat \eb_n^*)\big)  + \eta_n \widehat \omega_{n,j} \textrm{sgn}(\widehat \beta_{n,j}^*) . \nonumber
\end{align}
Because by Theorem \ref{thh} and Remark \ref{vit_el} the convergence rates of $\widehat \eb_n^*$ towards $\ebo$ and of $\widehat \el_n^* $ towards $\oo_p$ are of order  $n^{-1/3}$, let us consider  $\eb$ et $\el$  in the ball $\|\eb -\ebo\| + \| \el- \oo_p\| \leq C n^{-1/3}$. We do the Taylor expansion:
\begin{align*}
	\textbf{S}_{2}^*\big(\eb, \el(\eb)\big)&=  \bigg(\textbf{S}_2(\ebo, \oo_p)+(\eb-\ebo)^\top \frac{\partial  \textbf{S}_2(\ebo, \oo_p)}{\partial \eb}+(\el(\ebo)-\oo_p)^\top \frac{\partial \textbf{S}_2(\ebo,\oo_p)}{\partial \el}\bigg)  \\
	& \cdot \big( 1+o_\PP(1)\big)+\eta_n \big(\widehat \omega_{n,1}\textrm{sgn}(\beta_1), \cdots, \widehat \omega_{n,p} \textrm{sgn}(\beta_p)\big).
\end{align*}
On the other hand, we have $\textbf{S}_{2}(\ebo,\oo_p)=\oo_p$ and also $\frac{\partial \textbf{S}_{2}(\ebo,\oo_p)}{\partial \eb}=\oo_{p \times p}$ (with $\oo_{p \times p}$ a $p$-square matrix with all zero elements), which imply:
\begin{equation}
	\label{TS1}
	\textbf{S}_{2}^*\big(\eb, \el(\eb)\big)=  \big(\el(\ebo)- \oo_p\big)^\top \frac{\partial \textbf{S}_{2}(\ebo, \oo_p)}{\partial \el}  \big( 1+o_\PP(1)\big)+\eta_n   \big( \widehat \omega_{n,j}\textrm{sgn}(\beta_j)\big)_{1 \leqslant j \leqslant p}.
\end{equation}
The $j$-component $\textbf{S}_{2,j}(\eb, \el(\eb))$ of the vector  $  \textbf{S}_2(\eb, \el(\eb))$ is $2 n^{-1} \sum^n_{i=1} \frac{\el^\top  \frac{\partial \widehat \eg_i (\eb)}{\partial \beta_j}}{1+\el^\top \widehat \eg_i (\eb) }$. 
Then, by elementary calculations we get for any $j=1, \cdots,, p$:
\[
\frac{\partial \textbf{S}_{2,j}(\eb, \el(\eb))}{\partial \el^\top}=\frac{2}{n} \sum^n_{i=1} \bigg(\frac{\frac{\partial \widehat \eg_i (\eb)}{\partial \beta_j}}{ 1+\el^\top \widehat \eg_i (\eb)  }-\frac{ \el^\top \frac{\partial \widehat \eg_i (\eb)}{\partial \beta_j}\widehat \eg_i (\eb) }{\big(1+\el^\top \widehat \eg_i (\eb) \big)^2} \bigg),
\]
which is a  $p$-column  vector.  Note that the $p$-square matrix  $\frac{\partial \textbf{S}_2(\eb,\el(\eb))}{\partial \el}$ has the columns:
$ \bigg( \frac{\partial \textbf{S}_{2,j}(\eb,\el(\eb))}{\partial \el}\bigg)_{1 \leqslant j \leqslant p}$.\\
Then,  by relation (\ref{L33b}) we have $\frac{\partial \textbf{S}_2(\ebo,\oo_p)}{\partial \el}= 2n^{-1} \sum^n_{i=1} \bigg( \frac{\partial \widehat \eg_i(\eb)}{\partial \beta_j}\bigg)_{1 \leqslant j \leqslant p }=O_\PP(1)$. \\
  Let us consider an index $j \in \widehat {\cal A}_n \cap {\cal A}^c$. In this case, from Theorem \ref{Lemma 2.1 OA} we have $\widehat \beta_{n,j}=O_\PP(n^{-1/3})$, which implies: 
  \begin{equation}
  \label{TS0b1}
  \big| \eta_n \widehat \omega_{n,j}\big|=O_\PP\big(\eta_n n^{\gamma /3}\big).
  \end{equation}
By the supposition $n^{(\gamma +1)/3} \eta_n  {\underset{n \rightarrow \infty}{\longrightarrow}} \infty$ combines with $\el(\ebo)- \oo_p=O_\PP(n^{-1/3})$, we have that the penalty dominates in the right-hand side of  (\ref{TS1}). \\
We now return to  relation (\ref{TS0}), for which  relation (\ref{TS0b1}) together with the convergence rates of $\widehat \eb_n^*$ and of $\widehat \el_n^*$,   relations  (\ref{M1M2}), (\ref{L33b}), we obtain:
\begin{equation}
	\label{TS0b2}
	{S}_{2,j}\big(\widehat \eb_n^*, \el(\widehat \eb_n^*)\big)=O_\PP(\el_n^*)=O_\PP(n^{-1/3}).
\end{equation}
Since $n^{-1/3} \ll n^{\gamma/3} \eta_n$ for $n$ large enough, relations (\ref{TS0}), (\ref{TS0b1}), (\ref{TS0b2}) imply:
\begin{equation*}
	j \in \widehat {\cal A}^*_n \cap {\cal A}^c \overset{\eP}{\underset{n \rightarrow \infty}{\longrightarrow}} 0.
\end{equation*}
This relation gives us 
\begin{equation} 
	\label{kh}
	\lim_{n \rightarrow \infty} \PP[ \widehat {\cal A}^*_n \subseteq {\cal A} ]=1.
\end{equation}
On the other hand, by the consistency of $\widehat \eb_n^*$ we have that 
\begin{equation}
	\label{vb}
	\lim_{n \rightarrow \infty} \PP \big[{\cal A} \subseteq \widehat {\cal A}^*_n \big] =1.
\end{equation}
Then, relations (\ref{kh}) et (\ref{vb}) imply:
\[
\lim_{n \rightarrow \infty} \PP \big[{\cal A} = \widehat {\cal A}^*_n \big] =1.
\]
(\underline{ii) Asymptotic normality }\\
We denote  $\widehat \eb_n^*=\big( \widehat \eb_{n1}^*, \widehat \eb_{n2}^*\big)$, with $\widehat \eb_{n1}^*$ of dimension $q=|{\cal A}|$, given the sparsity property proved at \textit{(i)}. Again taking assertion (i)  into account and the convergence rate of  $\widehat \eb_n^*$, hereafter  we consider $\eb=(\eb_1,\oo)$, with $\| \eb_1 -\eb^0_1 \| =n^{-1/3}$ and $\ebo=(\eb^0_1, \oo_{p-q})$.   For the following square matrices of dimension $p \times p$ we have:
\begin{equation*}
	\left\{  
	\begin{split} 
		\frac{\partial \textbf{S}_{1}	(\eb, \oo_p)}{\partial \el}& = -2 n^{-1} \sum^n_{i=1} \widehat \eg_i (\eb) \widehat \eg_i (\eb)^\top, \\
	\frac{\partial \textbf{S}_{2}	(\eb, \oo_p)}{\partial \eb} &= \oo_{p \times p}.
	\end{split}
	\right. 
\end{equation*}
On the other hand, we have $\textbf{S}_{2}^*(\eb,\el)= \textbf{S}_2(\eb,\el)+\eta_n \big(\big(\widehat \omega_{n,j}  \textrm{sgn}(\beta_j)\big)_{j \in {\cal A}},\oo_{p-q}\big)$.\\
Moreover, $\widehat \el_n^*$ and $\widehat \eb_n^*$ satisfy the following relations, obtained by the Taylor expansions of (\ref{dRb}):
\begin{equation}
	\label{LR1}
	\left\{
	\begin{split} 
		\oo_p =&\textbf{S}_{1}^*(\widehat \eb_n^*,\widehat \el_n^*)=\bigg(\textbf{S}_{1}(\ebo,\oo_p) +\big(\widehat \eb_n^* - \ebo\big)^\top\frac{\partial \textbf{S}_1(\ebo , \oo_p)}{\partial \eb} \bigg. \\
			&  \qquad  + \bigg. (\widehat \el_n^*)^\top \frac{\partial \textbf{S}_1(\ebo, \oo_p)}{\partial \el} \bigg)  \big( 1+o_\PP(1) \big),\\
		\oo_p =&\textbf{S}_{2}^*(\widehat \eb_n^*,\widehat \el_n^*)=\bigg(\textbf{S}_{2}(\ebo,\oo_p) +\big(\widehat \eb_n^* - \ebo\big)^\top\frac{\partial \textbf{S}_2(\ebo , \oo_p)}{\partial \eb}  \bigg.   \\
		&  \qquad    \bigg. + (\widehat \el_n^*)^\top \frac{\partial \textbf{S}_2(\ebo, \oo_p)}{\partial \el}	+ \eta_n \big(\big(\widehat \omega_{n,j}  \textrm{sgn}(\beta_j)\big)_{j \in {\cal A}},\oo_{p-q}\big)	 \bigg) \big( 1+o_\PP(1) \big).
	\end{split}
	\right.
\end{equation}
Since $\textbf{S}_2(\ebo, \oo_p)=\oo_p$ and  $\frac{\partial \textbf{S}_2(\ebo , \oo_p)}{\partial \eb} =\oo_{p\times p}$, the second relation of (\ref{LR1})  becomes $\oo_p=\big(  (\widehat \el_n^*)^\top \frac{\partial \textbf{S}_2(\ebo, \oo_p)}{\partial \el}	+ \eta_n \big( \textbf{C}_q, \oo_{p-q}\big)	 \big) \big( 1+o_\PP(1) \big)$, with $\textbf{C}_q$ a constant $q$-vector, from where we deduce,
	\[
	\widehat \el^*_n=- \eta_n \big( \textbf{C}_q, \oo_{p-q}\big)    \bigg(  \frac{\partial \textbf{S}_2(\ebo, \oo_p)}{\partial \el}\bigg)^{-1} \big( 1+o_\PP(1) \big).
	\]
Then, we replace $\widehat \el_n^*$ in the first equation of (\ref{LR1}) and we obtain:
	\begin{equation}
	\begin{split}
		\oo_p=&\bigg\{ \textbf{S}_1(\ebo, \oo_p)+\big( \widehat \eb_n^* - \ebo\big)^\top \frac{\partial \textbf{S}_1(\ebo, \oo_p)}{\partial \eb}  - \eta_n \big( \textbf{C}_q, \oo_{p-q}\big)^\top  
	  \bigg(  \frac{\partial \textbf{S}_2(\ebo, \oo_p)}{\partial \el}\bigg)^{-1}\\
		& \qquad \qquad \qquad \cdot  \frac{\partial  \textbf{S}_1(\ebo, \oo_p)}{\partial \el}  \bigg\}  \big( 1+o_\PP(1) \big) .
		\end{split}
		\label{S1}
	\end{equation}
Furthermore, $\textbf{S}_1(\ebo, \oo_p)=2n^{-1} \sum^n_{i=1} \widehat \eg_i(\ebo)=O_\PP(n^{-1/2})$ by (\ref{L22}) and taking into account the supposition $n^{2/3}\eta_n {\underset{n \rightarrow \infty}{\longrightarrow}}  0$ we obtain $\eta_n  \big( \textbf{C}_q, \oo_{p-q}\big)=o(n^{-1/2}) $. These relations, together with  relation (\ref{S1})  imply
	\begin{equation*}
	\label{pap}
	\oo_p= \bigg( \textbf{S}_1(\ebo, \oo_p) +(\widehat \eb_n^* - \ebo)^\top  \frac{\partial \textbf{S}_1(\ebo, \oo_p)}{\partial \eb}\bigg)\big( 1+o_\PP(1) \big) .
	\end{equation*}
	On the other hand, by the LLN we have
	\[
	\frac{1}{2}	\frac{\partial \textbf{S}_1(\ebo, \oo_p)}{\partial \eb} = \frac{1}{n}  \sum^n_{i=1} \frac{\partial \widehat \eg_i (\ebo)}{\partial \eb}   \overset{\PP} {\underset{n \rightarrow \infty}{\longrightarrow}} \eE \bigg[ \frac{\partial \widehat \eg_i(\ebo)}{\partial \eb}\bigg].
	\]
From these relations we obtain $\widehat \eb_n^* - \ebo =O_\PP(n^{-1/2})$ and
	\begin{equation}
		\label{nbA}
	n^{1/2} \big(\widehat \eb_n^* -\ebo\big)_{\cal A} =\bigg( - \frac{1}{\sqrt{n}}  \sum^n_{i=1} \widehat \eg_i(\ebo)  \bigg)_{\cal A}^\top \bigg(\frac{1}{2}\frac{\partial \textbf{S}_1(\ebo, \oo_p)}{\partial \eb}\bigg)^{-1}_{\cal A}\big( 1+o_\PP(1) \big) .
	\end{equation}
Furthermore, from relation (\ref{eq1}) we have $\big( n^{-1/2}  \sum^n_{i=1} \widehat \eg_i(\ebo)  \big)_{\cal A}  \overset{\cal L} {\underset{n \rightarrow \infty}{\longrightarrow}} {\cal N}(\oo_q,\textbf{B}_{\cal A})  $, with the matrix $\textbf{B}$ defined by  relation (\ref{L5}).  
Then relation (\ref{nbA})  implies
	\[
	n^{1/2} \big(\widehat \eb_n^* -\ebo\big)_{\cal A}   \overset{\cal L} {\underset{n \rightarrow \infty}{\longrightarrow}} {\cal N}\big( \oo_{|{\cal A}|}, \eE \big[ \frac{\partial \widehat \eg_i(\ebo)}{\partial \eb}\big]^\top_{\cal A}   \textbf{B}_{\cal A}   \eE \big[ \frac{\partial \widehat \eg_i(\ebo)}{\partial \eb}\big]_{\cal A} \big),
	\]
i.e. the result obtained by   Theorem \ref{Th1 ZW} for MEL estimators corresponding to the smoothed expectile method, for non-zero coefficients.
	\hspace*{\fill}$\blacksquare$


\end{document}